\documentclass[12pt,a4paper]{article}
\usepackage{amsmath}
\usepackage{amsfonts}
\usepackage{amsthm}
\usepackage{hyperref}
\usepackage{graphicx}
\usepackage{todonotes}
\usepackage{color}

\newtheorem{theorem}{Theorem}
\newtheorem{lemma}[theorem]{Lemma}
\newtheorem{corollary}[theorem]{Corollary}
\newtheorem{proposition}[theorem]{Proposition}
\newtheorem{conjecture}[theorem]{Conjecture}

\newtheorem{myclaim}{Claim}[theorem]
\newtheorem{claim}{Claim}[theorem]

\newcommand{\NP}{\mathrm{NP}}
\newcommand{\XNLP}{\mathrm{XNLP}}
\def\poly{\operatorname{poly}}
\newcommand{\N}{{\bf N}} 
\newcommand{\Z}{{\bf Z}} 

\usepackage{array}
\usepackage{mdwmath}
\usepackage{mdwtab}
\usepackage{eqparbox}

\begin{document}
\title{Parameterized Problems Complete for \\Nondeterministic FPT time and Logarithmic Space%
\footnote{This paper contains the results reported in \cite{BodlaenderGNS21}, with the exception of results on reconfiguration, and
one result from \cite{Bodlaender20} (the reduction in the proof of Theorem~\ref{theorem:bandwidth}).}}

\author{Hans L. Bodlaender\thanks{Department of Information and Computing Sciences, Utrecht University, 
the Netherlands, h.l.bodlaender@uu.nl} 
\and
Carla Groenland\thanks{Faculty of Electrical Engineering, Mathematics and Computer Science, Technical University Delft, the Netherlands, c.e.groenland@tudelft.nl.
This research was done when Carla Groenland was associated with Utrecht University.}
\and
Jesper Nederlof\thanks{Department of Information and Computing Sciences, Utrecht University, the Netherlands, j.nederlof@uu.nl. The work of Jesper Nederlof was supported by the project CRACKNP that has received funding from the European
Research Council (ERC) under the European Union’s Horizon 2020 research and innovation programme (grant agreement No 853234).} 
\and
C\'{e}line Swennenhuis\thanks{Department of Mathematics and Computer Science, Eindhoven University of Technology, Eindhoven, The Netherlands, c.m.f.swennenhuis@tue.nl. The work of C\'{e}line Swennenhuis was supported by the Netherlands Organization for Scientific Research under
project no. 613.009.031b.}
}

\maketitle

\begin{abstract}
Let $\XNLP$ be the class of parameterized problems such that an instance of size $n$ with parameter $k$ can be solved nondeterministically in time $f(k)n^{O(1)}$ and space $f(k)\log(n)$ (for some computable function $f$). We give a wide variety of $\XNLP$-complete problems, such as {\sc List Coloring} and
    {\sc Precoloring Extension} with pathwidth as parameter,
    {\sc Scheduling of Jobs with Precedence Constraints}, with both number of machines and partial order width as parameter, {\sc Bandwidth} and variants of 
    {\sc Weighted CNF-Satisfiability}. In particular, this implies that all these problems are $\mathrm{W}[t]$-hard for all $t$. This also answers a long standing question on the parameterized complexity of the {\sc Bandwidth} problem.
\end{abstract}

%\begin{IEEEkeywords}
%Parameterized complexity; XNLP; Bandwidth; W-hierarchy
%\end{IEEEkeywords}

%%%%%%%%%%% Introduction %%%%%%%%%%%%%

\section{Introduction}
\label{section:introduction}
Already since the 1970's, an important paradigm in classical complexity theory has been that an increased number of alternations of existential and universal quantifiers increases the complexity of search problems: This led to the central definition of the polynomial hierarchy~\cite{Stockmeyer76}, whose study resulted in cornerstone results in complexity theory such as Toda's theorem and lower bounds for time/space tradeoffs for SAT~\cite{AroraBarak}.
In their foundational work in the early 1990s, Downey and Fellows introduced an analogue of this hierarchy for parameterized complexity theory, called the \emph{$\mathrm{W}$-hierarchy}. This hierarchy comprises of the complexity classes $\mathrm{FPT}$, the parameterized analogue of $\mathrm{P}$, $\mathrm{W}[1]$, the parameterized analogue of $\NP$, and the classes $\mathrm{W}[2]$, \ldots, $\mathrm{W[P]}$, $\mathrm{XP}$ (see e.g.~\cite{DowneyF95,DowneyF95a,DowneyF99}). 

While in the polynomial hierarchy only the classes with no quantifier alternation (i.e. $\mathrm{P}, \NP$
and $\mathrm{coNP}$) are prominent, many natural parameterized problems are known to be hard or even complete for $\mathrm{W}[i]$ for some $i \geq 1$. 
Thus, the $\mathrm{W}$-hierarchy substantially differentiates the complexity of hard parameterized problems. And such a differentiation has applications outside parameterized complexity as well: For example, for problems in $\mathrm{W}[1]$ we can typically improve over brute-force enumeration algorithms, while for problems in $\mathrm{W}[2]$ we can prove lower bounds under the Strong Exponential Time Hypothesis excluding such improvements (see e.g. the discussion in~\cite{AbboudLW14}).\footnote{For example the na\"ive algorithm $O(n^{k+1})$ time algorithm for finding cliques on $k$ vertices on $n$-vertex graphs can be improved to run in $n^{0.8 k}$ time, but similar run times for {\sc Dominating Set} refute the Strong Exponential Time Hypothesis.}

For many problems,
completeness for a class is known, e.g., {\sc Clique} is $\mathrm{W}[1]$-complete \cite{DowneyF95a} and
{\sc Dominating Set} is $\mathrm{W}[2]$-complete \cite{DowneyF95}. 
However, there are also several problems known to
be hard for $\mathrm{W}[1]$, $\mathrm{W}[2]$, or even for $\mathrm{W}[t]$ for all positive integers $t$, but which
are not known to be in the class $\mathrm{W[P]}$; in many cases, only membership in $\mathrm{XP}$ was known.
For such problems, it is an intriguing question to establish their exact position within the $\mathrm{W}$-hierarchy as it can be expected to shed light on their complexity similarly as it did for the previous problems.

One example of such a problem is the {\sc Bandwidth} problem. It has been known to be hard for all classes $\mathrm{W}[t]$
since 1994 \cite{BodlaenderFH94}.
Already in the midst of the 1990s, Hallett argued that it is unlikely that {\sc Bandwidth} belongs to $\mathrm{W[P]}$,
see the discussion by Fellows and Rosamond in \cite{FellowsR20}. The argument intuitively boils down to the following: {\sc Bandwidth} `seems' to need certificates with $\Omega(n)$ bits, while problems
in $\mathrm{W[P]}$ have certificates with $O(f(k)\log n)$ bits. A similar situation applies to several other
$\mathrm{W}[1]$-hard problems. 

A (largely overlooked) 
breakthrough was made a few years ago by Elberfeld, Stockhusen and Tantau~\cite{ElberfeldST15},
who studied several classes of parameterized problems, including a class which they
called $\operatorname{N}[f \poly, f \log]$. This class is defined as the set parameterized problems that can be solved with a non-deterministic algorithm with simultaneously, the running
time bounded by $f(k)n^c$ 
and the space usage bounded by $f(k)\log n$, with $k$ the parameter, $n$ the
input size, $c$ a constant, and $f$ a computable function. For easier future reference, we denote this class by
$\XNLP$. Elberfeld et al.~\cite{ElberfeldST15} showed that a number of problems are complete for this class,
including the {\sc Longest Common Substring} problem. Since 1995, {\sc Longest Common Substring}
is known to be hard for all 
$\mathrm{W}[t]$~\cite{BodlaenderDFHW95}, but its precise parameterized
complexity was unknown until the result by Elberfeld et al.~\cite{ElberfeldST15}.

\paragraph{Our contribution} We show that the class $\XNLP$ (i.e., $\operatorname{N}[f \poly, f \log]$) can play an important role in establishing
the parameterized complexity of a large collection of well studied problems, ranging from abstract
problems on different types of automata (see e.g.~\cite{ElberfeldST15} or later in this paper),
logic, graph theory, scheduling, and more. 
In this paper, we give a number of different examples of problems
that are complete for $\XNLP$. These include {\sc Bandwidth}, thus indirectly answering a question
that was posed over 25 years ago. 

\begin{table}[htb]
    \centering
    \begin{tabular}{|c|c|}
    \hline
       Problem & Source \\ \hline
        {\sc Longest Common Subsequence} $+$ & \cite{ElberfeldST15} \\
        {\sc Timed Non-determ.~Cellular Automaton} $+$ & \cite{ElberfeldST15}, see Subsection~\ref{subsection:cellularautomata}\\
        {\sc Chained CNF-Satisfiability} $+$ & Subsection~\ref{subsection:satisfiability} \\
        {\sc Chained Multicolored Clique} & Subsection~\ref{subsection:chainedmcclique} \\
        {\sc Binary CSP} pw $+$ & Subsection~\ref{subsection:binarycsp} \\
        {\sc Accepting NNCCM} & Subsection~\ref{subsection:nncc} \\
        {\sc List Coloring} pw $+$ & Subsection~\ref{subsection:listcoloringpathwidth} \\
        {\sc Log-Pathwidth DS, IS} & Subsection~\ref{subsection:logpathwidth} \\
        {\sc Scheduling with precedence constaints} & Subsection~\ref{subsection:scheduling} \\
        {\sc Uniform Emulation of Weighted Paths} & Subsection~\ref{subsection:uepath} \\
        {\sc Bandwidth} & Subsection~\ref{subsection:bandwidth}\\
        {\sc Acyclic Finite State Automata Intersection} & \cite{Wehar16}; Subsection~\ref{section:fsaintersection} \\
        \hline
    \end{tabular}
    \caption{An overview of $\XNLP$-complete problems is given. For problems marked with $+$, the source also gives an $\XNLP$-hardness or completeness proofs for variants of the stated problem. We use the abbreviations CL = Clique, IS = Independent Set, DS = Dominating Set, pw = parameterized by pathwidth.}
    \label{table:xnlphard}
\end{table}

In Table~\ref{table:xnlphard}, we list the problems shown to be $\XNLP$-complete in either this paper
or by Elberfeld et al.~\cite{ElberfeldST15}. 

\begin{figure}[htb]
    \centering
    \includegraphics[width=\textwidth]{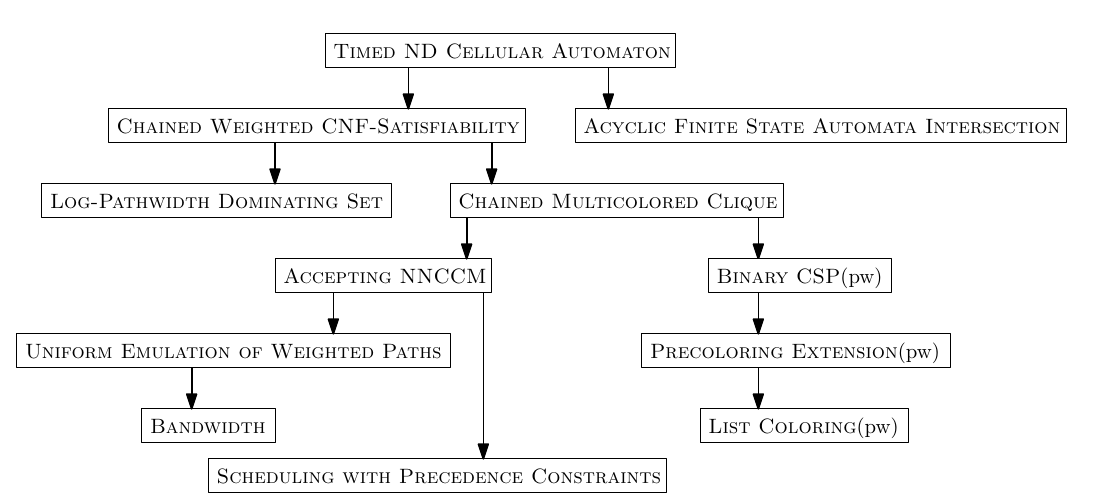}
    \caption{Reductions between $\XNLP$-hard problems from this paper. Several variants of problems are not shown.} 
    \label{figure:transformation}
\end{figure}

Figure~\ref{figure:transformation} shows for the problems from which problem the reduction starts to show $\XNLP$-hardness.

\medskip 

Often, membership in $\XNLP$ can be seen by looking at the algorithm that establishes membership in $\mathrm{XP}$.
Many
problems in $\XNLP$ typically have a dynamic programming algorithm that sequentially builds tables, with
each individual table entry expressible with $O(f(k) \log n)$ bits. We then get membership in $\XNLP$ by
instead of tabulating all entries of a table, guessing one entry of the next table --- this step
resembles the text-book transformation between a deterministic and non-deterministic finite automaton.

Interestingly, hardness for the class $\XNLP$ also has consequences for the use of
memory of deterministic parameterized algorithms. Pilipczuk and Wrochna~\cite{PilipczukW18}
conjecture that {\sc Longest Common Subsequence} (variant 1) has no $\mathrm{XP}$ algorithm that runs in
$f(k) n^c$ space, for a computable function $f$ and constant $c$; if this conjecture holds, then
no $\XNLP$-hard problem has such an algorithm. See Section~\ref{section:conclusions} for more details.

When a problem is $\XNLP$-hard, it is also hard for each class $\mathrm{W}[t]$ (see Lemma~\ref{lemma:xnlpwt}). Thus,
$\XNLP$-hardness proofs are also a tool to show hardness for $\mathrm{W}[t]$ for all $t$. In this
sense, our results strengthen existing results from the literature: for example, {\sc List Coloring}
and {\sc Precoloring Extension} parameterized by pathwidth (or treewidth)
were known to be $\mathrm{W}[1]$-hard \cite{FellowsFLRSST11}, and {\sc Precedence Constraint $K$-processor Scheduling} parameterized by the number of processors $K$ was known to be $\mathrm{W}[2]$-hard
\cite{BodlaenderF95}. Our $\XNLP$-hardness proofs imply hardness for $\mathrm{W}[t]$ for all $t$. Moreover, our $\XNLP$-hardness proofs are often simpler than the existing proofs that problems are hard for $\mathrm{W}[t]$ for all $t$. 

Related to the class $\XNLP$ is the class $\mathrm{XNL}$: the parameterized problems that can be solved by a 
nondeterministic algorithm that uses $f(k) \log n$ space. There is no explicit time bound,
but we can freely add a time bound of $2^{f(k)\log n}$, and thus $\mathrm{XNL}$ is a subset of $\mathrm{XP}$. $\mathrm{XNL}$ can be
seen as the parameterized counterpart of $\mathrm{NL}$. Amongst others, $\mathrm{XNL}$ was studied by Chen et al.~\cite{ChenF03}, who showed that {\sc Compact Turing Machine Computation} is complete
for $\mathrm{XNL}$.

Hardness for a class is always defined with respect to a class of reductions. In our proofs, we
use parameterized logspace reductions (or, in short, pl-reductions). A brief discussion of other
reductions can be found in Subsection~\ref{section:conclusion-reductions}.

\paragraph{Subsequent work}
After this paper appeared, several other problems were shown to be XNLP-complete, and related complexity classes
were defined, with their own complete problems. We briefly discuss these results here.

Recent XNLP-completeness results that build upon our work include the following:
\begin{itemize}
    \item several graph problems with a linear structure, including problems with pathwidth, linear cliquewidth, and linear mim-width as parameter \cite{BodlaenderGJJL22};
    \item $b$-coloring with pathwidth as parameter \cite{JaffkeLS22};
    \item several problems related 
to flow, with pathwidth as parameter \cite{BodlaenderCW22};
\item integral 2-commodity flow with pathwidth as parameter \cite{BodlaenderMOPvL23}.
\end{itemize}
In~\cite{reconfiguration}, it was shown that reconfiguration of
dominating sets and of independent sets, with the sizes of these sets as parameter is XNLP-complete, when the number of reconfiguration steps is given in unary. In contrast, when the number of steps is respectively a parameter, given in binary, or unspecified, the problems become complete for $W[2]$ or $W[1]$, for XNL, and for XL.

The class XALP was introduced by Bodlaender, Groenland, Jacob, Pilipczuk and Pilipzcuk
\cite{BodlaenderGJPP22} as an analogue to XNLP for `tree-structured' problems. Natural problems with a `tree structure', such as \textsc{List Coloring} parameterized by treewidth or \textsc{Max Cut} parameterized by clique-width were shown to be XALP-complete. Determining the tree partition width (or strong treewidth) of a graph was also shown to be XALP-complete in  \cite{BodlaenderGJ22} and the \textsc{Perfect Phylogeny} problem was shown to be XALP-complete by de Vlas~\cite{deVlas2023}. A probabilistic variant of XNLP with
a complete problem related to Bayesian networks was introduced in 
\cite{BodlaenderDK22}. Bodlaender, Groenland, and Pilipczuk~\cite{BodlaenderGP23} introduced the class XSLP, and showed
some natural problems with treedepth as parameter to be complete for that class.

\paragraph{Paper overview.} 
In Section~\ref{section:definitions}, we give a number of preliminary definitions and results. In Section~\ref{section:buildingblocks} we introduce three new problems that are $\XNLP$-complete. In Section~\ref{section:applications} we then use these problems as building blocks, to prove other problems to be either $\XNLP$-complete or $\XNLP$-hard. For each of the problems, its background and a short literature review specific to it will be given inside its relevant subsection. 
Final comments and open problems are
given in Section~\ref{section:conclusions}.

%%%%%%%%%%%%% Preliminaries %%%%%%%%%%

\section{Preliminaries}
\label{section:definitions}
In this section we formally define the class $\XNLP$ and give some preliminary results.

The section is organized as follows: first we introduce some basic notions in Subsection~\ref{subsec:basicnotions}, next we formally define the class $\XNLP$ in Subsection~\ref{subsec:defXNLP}. In Subsection~\ref{subsec:reductions} we then introduce the type of reductions that will be used in this paper and in Subsection~\ref{subsec:previousknown} we list some preliminary results. Subsection~\ref{subsection:cellularautomata} ends the section with a discussion of Cellular automata, for which Elberfeld et al.~\cite{ElberfeldST15} already established it was $\XNLP$-complete. From this problem we will (indirectly) derive the $\XNLP$-hardness for all other $\XNLP$-hard problems in this paper; containment in $\XNLP$ will always be argued more directly.

\subsection{Basic notions}
\label{subsec:basicnotions}
We assume the reader to be familiar with a number of notions from complexity theory, parameterized algorithms, and graph theory. A few of these are reviewed below, along with some new and less well-known notions.

We use interval notation for sets of integers, i.e., $[a,b] = \{ i\in \mathbb{Z} \mid a\leq i\leq b\}$. All logarithms in this paper have base $2$. $\N$ denotes the set of the natural numbers $\{0, 1, 2, \ldots \}$, and $\Z^+$ denotes the set of the positive natural numbers $\{1,2, \ldots\}$.

\subsection{Definition of the class $\XNLP$}
\label{subsec:defXNLP}

In this paper, we study parameterized decision problems, which are subsets of $\Sigma^\ast \times \N$, for a finite alphabet $\Sigma$.
The following notation is used, also by e.g. \cite{ElberfeldST15}, to denote classes of (non-)deterministic parameterized decision problems with a bound on the used time and space. Here, we use the following notations: $n$ for the input size;  $\poly$ for a polynomial function in the input size (i.e. $n^{O(1)}$); $\log$ for a logarithmic function in the input size (i.e. $O(\log n)$); $f$ for a computable
function of the parameter; $\infty$ if there is no a priory bound for the resource.

On top of the running time of an algorithm we will also study the \emph{space usage}. Informally, an algorithm has (random) read-only access to an input tape and random write-only access to an output tape. Additionally, it has both random read and write access to a working tape, but the length of the working tape equals the space usage.

Let $\operatorname{D}[t,s]$ denote the class of parameterized decision problems that can be solved by a deterministic algorithm in $t$ time and $s$ space and let $\operatorname{N}[t,s]$ be analogously defined for non-deterministic algorithms. Thus, $\mathrm{FPT}$ can be denoted by $\operatorname{D}[f \poly, \infty]$; we can denote
$\mathrm{XP}$ by $\operatorname{D}[n^f,\infty]$, $\NP$ by $\operatorname{N}[\poly,\infty]$, $L$ by $\operatorname{D}[\infty, \log ]$, etcetera.

A special role in this paper is played by the class $\operatorname{N}[f \poly, f \log]$: the parameterized decision problems that can be solved by a non-deterministic algorithm that simultaneously uses at most $f(k) n^c$
time and at most 
$f(k) \log n$ space, on an input $(x,k)$, where $x$ can be encoded with $n$ bits, $f$ a computable function, and $c$ a constant. Because of the special role of this class, we use the shorter notation
$\XNLP$.

$\XNLP$ is a subclass of the class $\mathrm{XNL}$, which was studied by Chen et al.~\cite{ChenF03}. $\mathrm{XNL}$ is the
class of problems solvable with a non-deterministic algorithm
in $f(k)\log n$ space ($f$, $k$, $n$ as above), i.e, $\mathrm{XNL}$ is the class $\operatorname{N}[\infty, f \log]$.

We assume the reader to be familiar with notions from parameterized complexity, such as $\mathrm{XP}$, $\mathrm{W}[1]$,
$\mathrm{W}[2]$, \ldots, $\mathrm{W[P]}$ (see e.g.  \cite{DowneyF99}). For classes of parameterized problems, we can often make a distinction between non-uniform (a separate algorithm for each parameter value), and
uniform. Throughout this paper, we look at the uniform variant of the classes, but we also will assume that functions $f$ of the parameter in time and resource bounds are computable --- this is called {\em strongly uniform} by
Downey and Fellows~\cite{DowneyF99}.

\subsection{Reductions}
\label{subsec:reductions}
Hardness for a class is defined in terms of reductions. We mainly use parameterized logspace reductions (defined below), which are a special case of fixed parameter tractable reductions. Both are defined below; the definitions
are based upon the formulations in \cite{ElberfeldST15}. Two other types of reductions are briefly discussed in the conclusion (Section~\ref{section:conclusion-reductions}.)

\begin{itemize}
    \item A {\em parameterized reduction} from a parameterized problem $Q_1 \subseteq \Sigma_1^\ast \times \N$ to a parameterized problem $Q_2 \subseteq \Sigma_2^\ast \times \N$ is a function
    $f : \Sigma_1^\ast \times \N \rightarrow \Sigma_2^\ast \times \N$, such that the following holds.
    \begin{enumerate}
        \item For all $(x,k) \in \Sigma_1^{\ast} \times \N$, $(x,k)\in Q_1$ if and only if $f((x,k)) \in Q_2$.
        \item There is a computable function $g$, such that for all $(x,k) \in \Sigma_1^\ast \times \N$, if $f((x,k)) = (y,k')$, then $k' \leq g(k)$.
    \end{enumerate}
    \item A {\em parameterized logspace reduction} or {\em pl-reduction} is a parameterized reduction
    for which there is an algorithm that computes $f((x,k))$ in space $O(g(k) + \log n)$, with $g$ a computable
    function and $n=|x|$ the number of bits to denote $x$. 
     \item A {\em fixed parameter tractable reduction} or {\em fpt-reduction} is a parameterized reduction
    for which there is an algorithm that computes $f((x,k))$ in time $O(g(k) n^c)$, with $g$ a computable
    function, $n=|x|$ the number of bits to denote $x$ and $c$ a constant.
\end{itemize}

A deterministic algorithm that uses $O(g(k)+ \log n)$ space necessarily runs in $2^{O(g(k)+ \log n)}$ time which is FPT in $k$. Thus, if there is a pl-reduction from problem $A$ to problem $B$ and $B$ is in $\XNLP$, then problem $A$ is in $\XNLP$ as well.

In the remainder of the paper, 
completeness for $\XNLP$ is with respect
to pl-reductions.\footnote{Note we could also have defined completeness for $\XNLP$ with respect to slightly more general reductions allowing algorithms using $O(g(k)\log n)$ space and time FPT in $k$, since such reductions still preserve containment in $\XNLP$. We use the more strict variant since all our reductions use only $O(g(k)\log n)$ space.}

\subsection{Preliminary results on $\XNLP$}
\label{subsec:previousknown}
We give some easy observations that relate $\XNLP$ to other notions from parameterized complexity. 
The following easy observation can be seen as a special case of the fact that $\operatorname{N}[\infty,S(n)] \subseteq \operatorname{D}[2^{S(n)},\infty]$, see \cite[Theorem 4.3]{AroraBarak}.

\begin{lemma}
$\XNLP$ is a subset of $\mathrm{XP}$.
\end{lemma}

\begin{proof}
Using standard techniques, we can transform the non-deterministic algorithm to a deterministic algorithm that employs dynamic programming: tabulate all reachable configurations of the machine --- a configuration 
is a tuple, consisting of the contents of the work tape, the state of the machine, and the position of the two headers. From a configuration, we can compute all configurations that can be reached in one step, and thus we can check if a configuration that has an accepting state can be reached. 

The number of such configurations is bounded by the product of a single exponential of the size of the work tape (i.e., at most $2^{f(k) \log n} = n^{f(k)}$ for some computable function $f$), the constant number of states of the machine, and the $O(f(k) \log n) \cdot n$ number of possible pairs of locations of the heads, and thus bounded by a function of the form $n^{g(k)}$ with $g$ a computable function.
\end{proof}

\begin{lemma}
If a parameterized problem $Q$ is $\XNLP$-hard, then it is hard for each class $\mathrm{W}[t]$ for all $t\in \Z^+$. 
\label{lemma:xnlpwt}
\end{lemma}

\begin{proof}
Observe that the $W[t]$-complete problem {\sc Weighted $t$-Normalized Satisfiability}  belongs to $\XNLP$.
(In {\sc Weighted $t$-Normalized Satisfiability}, we have a Boolean formula with parenthesis-depth $t$ and
ask if we can satisfy it by setting exactly $k$ variables to true and all others to false; we
can non-deterministically guess which of the $k$ Boolean variables are true; verifying whether this
setting satisfies the formula can be done with $O(t+ k\log n)$ bits of space, see e.g.~\cite{DowneyF99}.)

Each problem in $W[t]$ has an fpt-reduction to {\sc Weighted $t$-Normalized Satisfiability}, and
the latter has a pl-reduction (which is also an fpt-reduction) to any $\XNLP$-hard problem $Q$. The transitivity of
fpt-reductions implies that $Q$ is then also hard for $W[t]$.
\end{proof}

\begin{lemma}[Chen et al.~\cite{ChenF03}]
If $\mathrm{NL}\neq \mathrm{P}$, then there are parameterized problems in $\mathrm{FPT}$ that do not belong to $\mathrm{XNL}$ (and hence also not to $\XNLP$).
\end{lemma}

\begin{proof}
Take a problem $Q$ that belongs to P, but not to NL. Consider the parameterized problem $Q'$
with $(x,k)\in Q'$ if and only if $x \in Q$. (We just ignore the parameter part of the input.) Then $Q'$ belongs to FPT, since the polynomial time algorithm for $Q$ also solves $Q'$. If $Q'$ is in XNL,
then there is an algorithm that solves $Q$ in (non-deterministic) logarithmic space, a contradiction. So $Q'$ belongs to FPT but not to XNL.
\end{proof}
Chen et al.~\cite{ChenF03} introduce the following problem.

\newpage
%% Page break forced here for layout purposes! Remove when updating the paper
%%

\begin{verse}
{\sc CNTMC (Compact Nondeterministic Turing Machine Computation)}\\
{\bf Input}: the encoding of a non-deterministic Turing Machine $M$; the encoding of a string $x$ over the alphabet of the machine.\\
{\bf Parameter}: $k$.\\
{\bf Question}: Is there an accepting computation of $M$ on input $x$ that visits at most $k$ cells of the work tape?
\end{verse} 

\begin{theorem}[Chen et al.~\cite{ChenF03}]
{\sc CNTMC} is $\mathrm{XNL}$-complete under pl-reductions.
\label{theorem:chen}
\end{theorem}

It is possible to show $\XNLP$-completeness  for a `timed' variant of this problem.

\medskip \begin{verse}
{\sc Timed CNTMC}\\
{\bf Input}: the encoding of a non-deterministic Turing Machine $M$; the encoding of a string $x$ over the alphabet of the machine; an integer $T$ given in unary.\\
{\bf Parameter}: $k$.\\
{\bf Question}: Is there an accepting computation of $M$ on input $x$ that visits at most $k$ cells of the work tape and uses at most $T$ time?
\end{verse} \medskip
The fact that the time that the machine uses is given in unary, is needed to show membership in $\XNLP$.
\begin{theorem}
{\sc Timed CNTMC} is $\XNLP$-complete.
\end{theorem}
We state the result without proof, as the proof is similar to the proof of Theorem~\ref{theorem:chen} from \cite{ChenF03},
and we do not build upon the result. We instead start with a problem on cellular automata which was shown to be complete for $\XNLP$ by Elberfeld et al.~\cite{ElberfeldST15}. We discuss
this problem in the next subsection. Elberfeld et al.~\cite{ElberfeldST15}
show a number of other problems to be $\XNLP$-complete, including a timed version of the acceptance of multihead automata, and the {\sc Longest Common Subsequence} problem, parameterized by the number of strings. The latter result is discussed in the Conclusion, Section~\ref{section:spaceefficiency}.

%%%%%%%%%%%% Cellular automata %%%%%%%%%%%%%%

\subsection{Cellular automata}
\label{subsection:cellularautomata}

In this subsection, we discuss one of the results by Elberfeld et al.~\cite{ElberfeldST15}. Amongst the
problems that are shown to be complete for $\XNLP$ by Elberfeld et al.~\cite{ElberfeldST15}, of central
importance to us is the {\sc Timed Non-deterministic Cellular Automaton} problem. 
We use the hardness of this problem to show the hardness of {\sc Chained CNF-Satisfiability}
in Subsection~\ref{subsection:satisfiability}.

In this subsection, we describe the 
{\sc Timed Non-deterministic Cellular Automaton} problem, and a variant. We are
given a linear cellular automaton, a time bound $t$ given in unary, and a starting configuration for the automaton, and ask if after $t$ time steps, at least one cell is in an accepting state.

More precisely, we have a set of states $S$, and subset of accepting states $A \subseteq S$. We assume there are two special states $s_L$ and $s_R$ which
are used for the leftmost and rightmost cell. A {\em configuration} is a function $c: \{1, \ldots, q\}\to S$,
with $c(1)=s_L$, $c(q)=s_R$ and for $i\in [2,q-1]$, $c(i)\in S \setminus \{s_L,s_R\}$. We say that
we have $q$ cells, and in configuration $c$, cell $i$ has state $c(i)$.
The machine is further described by a collection of 4-tuples $\cal T$ in $S \times (S\setminus \{s_L,s_R\}) \times S \times (S\setminus \{s_L,s_R\})$.
At each time step, each cell $i\in [2,q]$ reads the 3-tuple $(s_1, s_2, s_3)$ of states given by the current states of the cells $i-1,~ i$ and $i+1$ (in that order). 
If there is a cell $i=1,\ldots,q$ with $s_i \in A$, then the machine accepts.
If there is no 4-tuple of the form $(s_1,s_2,s_3,s_4)$ for some
$s_4\in S$, then the machine halts and rejects; otherwise, the cell selects an $s_4\in S$ with
$(s_1,s_2,s_3,s_4)\in {\cal T}$ and moves in this time step to state $s_4$. (In a non-deterministic machine, there can be multiple such states $s_4$ and a non-deterministic step is done. For a deterministic
cellular automaton, for each 3-tuple $(s_1,s_2,s_3)$ there is at most one 4-tuple $(s_1,s_2,s_3,s_4)\in {\cal T}$.) Note that the leftmost and rightmost cell never change state: their states are used to mark the ends of the tape of the automaton. 

\medskip \begin{verse}
{\sc Timed Non-deterministic Cellular Automaton}\\
{\bf Input}: Cellular automaton with set of states $S$ and set of transitions $\cal T$; configuration $c$ on $q$ cells; integer in unary $t$; subset $A\subseteq S$ of accepting states. \\
{\bf Parameter:} $q$.\\
{\bf Question:} Is there an execution of the machine for exactly $t$ time steps with initial configuration $c$, such that at time step $t$ 
at least one cell of the automaton is in $A$?
\end{verse} \medskip
We will build on the following result.
\begin{theorem}[Elberfeld et al.~\cite{ElberfeldST15}]
{\sc Timed Non-deterministic Cellular Automaton} is $\XNLP$-complete.
\end{theorem}

We recall that the class, denoted by $\XNLP$ in the current paper, is called \linebreak $\operatorname{N}[f \poly, f \log ]$ in
\cite{ElberfeldST15}.

Elberfeld et al.~\cite{ElberfeldST15} state that asking that all cells are in an accepting state does not make a difference, i.e., if we modify the {\sc Timed Non-deterministic Cellular Automaton} problem by asking if all cells are in an accepting state at time $t$, then we also have an $\XNLP$-complete problem.

We also discuss a variant that can possibly be useful as another starting point for reductions.

\medskip \begin{verse}
{\sc Timed Non-halting Non-deterministic Cellular Automaton}\\
{\bf Input}: Cellular automaton with set of states $S$ and set of transitions $\cal T$; configuration $c$ on $q$ cells; integer in unary $t$; subset  $A\subseteq S$ of accepting states. \\
{\bf Parameter:} $q$. \\
{\bf Question:} Is there an execution of the machine for exactly $t$ steps with initial configuration $c$, such that the machine does not halt before time $t$?
\end{verse} \medskip

\begin{corollary}
{\sc Timed Non-halting Non-deterministic Cellular Automaton} is $\XNLP$-complete.
\end{corollary}

\begin{proof}
Membership in $\XNLP$ follows in the same way as for {\sc Timed Non-deter\-mi\-nis\-tic Cellular Automaton}, see
\cite{ElberfeldST15}; observe that we can store a configuration using $\lceil q \log |S|\rceil$ bits.

Hardness follows by modifying the automaton as follows. We take an automaton that accepts, if and only if
at time $t$ all cells are in an accepting state. Now, we enlarge the set of states as follows:
for each time step $t' \in [0,t]$, and each state $s\in S\setminus \{s_L,s_R\}$, we create
a state $s^{t'}$. The initial configuration $c$ is modified to $c'$ by setting $c'(i)= s^0$ for $i\in [2,q-1]$
when $c(i)=s$. 
We enlarge the set of transitions as follows. For each $t' \in [0,t-1]$
and $(s_1,s_2,s_3,s_4)\in {\cal T}$, we create a transition $(s_1^{t'},s_2^{t'},s_3^{t'},s_4^{t'+1})$ in the new set 
of transitions. In this way, each state of the machine also codes the time: at time $t'$ all cells except the first and last will have a state
of the form $s^{t'}$.

We run the machine for one additional step, i.e., we increase $t$ by one. We create one additional accepting state
$s_a$. For each accepting state
$s\in A$, we make transitions $(x, s^t, y, s_a)$ for all possible values $x$ and $y$ can take. When $s\not\in A$, then there are no $x,y,z$ for which there is a transition of the form $(x, s^t, y, z)$.
This ensures that a cell has a possible transition at time $t$ if and only if it is in an accepting state. In particular, when all states are accepting, all cells have a possible transition at time $t$; if there is a state that is not accepting at time $t$, then the machine halts.
\end{proof}

\subsection{Pathwidth, bandwidth, and cutwidth}
A \textit{path decomposition} of a graph $G=(V,E)$ is a sequence $(X_1, X_2, \ldots, X_r)$ of subsets of $V$  with 
the following properties.
\begin{enumerate}
    \item $\bigcup_{1\leq i\leq r} X_i = V$.
    \item For all $\{v,w\}\in E$, there is an $i\in I$ with $v,w\in X_i$.
    \item For all $1\leq i_0<i_1<i_2 \leq r$, $X_{i_0}\cap X_{i_2} \subseteq X_{i_1}$.
\end{enumerate}
The {\em width} of a path decomposition $(X_1, X_2, \ldots, X_r)$ equals $\max_{1\leq i\leq r} |X_i|-1$, and the {\em pathwidth} of a graph $G$ is the minimum width of a path decomposition of $G$.

When considering the parameter pathwidth, we will assume that a path decomposition of width at most $k$ is given as part of the input.
It currently is an open problem whether such a path decomposition can be found with a non-deterministic algorithm using 
logarithmic space and `fpt' time.
Kintali and Munteanu~\cite{KintaliMunteanuPathwidthL} show that for each fixed $k$, determining if the pathwidth is at most $k$,
and if so, finding a path decomposition of width at most $k$ belongs to $\mathrm{L}$.
As a subroutine, this uses a related earlier result from 
Elberfeld et al.~\cite{ElberfeldJT10}, who showed that for each fixed $k$, determining if the treewidth is at most $k$,
and if so, finding a tree decomposition of width at most $k$ belongs to $\mathrm{L}$.

A \emph{linear ordering} of a graph $G=(V,E)$
is a bijection $f: V \rightarrow [1,|V|]$. 
The \emph{bandwidth} of a linear ordering $f$ of $G=(V,E)$ is $\max_{\{v,w\}\in E} |f(v)-f(w)$.
The \emph{cutwidth} of a linear ordering $f$ of $G=(V,E)$ is 
$\max_{i\in [1,|V|-1]} |\{\{v,w\}\in E ~|~ f(v)\leq i<f(w)\}$. The \emph{bandwidth}, respectively
\emph{cutwidth} of a graph $G$ is the minimum
bandwidth, respectively cutwidth over all linear
orderings of $G$.

\section{Building Blocks}
\label{section:buildingblocks}
%%%%%%%%%%% Chained CNF-Satisfiability %%%%%%%%%%%%
In this section, we introduce three new problems and prove that they are $\XNLP$-complete, namely {\sc Chained CNF-Satisfiability}, {\sc Chained Multicolored Clique} and {\sc Accepting NNCCM}. These problems are called building blocks, as their main use is proving $\XNLP$-hardness for many other problems (see Figure~\ref{figure:transformation}). 

\subsection{{\sc Chained CNF-Satisfiability}}
\label{subsection:satisfiability}
In this subsection we give a useful starting point for our transformations: a variation of {\sc Satisfiability} which we call
{\sc Chained Weighted CNF-Satisfiability}. The problem can be seen as a generalization of
the $\mathrm{W}[1]$-hard problem {\sc Weighted CNF-Satis\-fiability} \cite{DowneyF95}.

\medskip \begin{verse}
{\sc Chained Weighted CNF-Satisfiability}\\
{\bf Input}: $r$ disjoint sets of Boolean variables $X_1, X_2, \ldots X_r$, each of size $q$; integer $k\in \N$; Boolean formulas
 $F_1$, $F_2$, \ldots, $F_{r-1}$, where each $F_i$ is an expression in conjunctive
normal form on variables $X_i \cup X_{i+1}$. \\
{\bf Parameter}: $k$. \\
{\bf Question}: Is it possible to satisfy the formula $F_1 \wedge F_2 \wedge \cdots \wedge F_r$ by
setting exactly $k$ variables to true from each set $X_i$  and all others to false?
\end{verse} \medskip 

Our main result in this subsection is the following. We will also prove a number of variations to be $\XNLP$-complete later in this subsection.

\begin{theorem}\label{theorem:chainedweightedcnfsat}
{\sc Chained Weighted CNF-Satisfiability} is $\XNLP$-complete.
\end{theorem}
\begin{proof}
Membership in $\XNLP$ is easy to see. Indeed, for $i$ from $1$ to $r$, we  non-deterministically guess which variables in each $X_i$ are true, and keep the indexes of the true
variables in memory for the two sets $X_i$, $X_{i+1}$. Verifying $F_i(X_i,X_{i+1})$ can easily be done in $O(k + \log n)$ space and linear time.

To show hardness, we transform from {\sc Timed Non-deterministic Cellular Automaton}.

For each time step $t'$, each cell $r\in [1,q]$, and each state $s\in S$, we have a Boolean variable $x_{t',r,s}$ with $x_{t',r,s}$ denoting whether
the $r$th cell of the automaton at time $t'$ is in state $s$. For each time step $t'\in [1,t-1]$, each cell $r\in [2,q]$, and each transition $z\in {\cal T}$ we have a variable $y_{t',r,z}$ that expresses that cell $r$ uses transition $z$ at time $t'$.

We will build a Boolean expression and partitions of the variables, such that the expression is satisfiable by setting exactly $k$ variables to true from each set in the partition if and only if the machine can reach time step $t$ with at least one cell in accepting state starting from the initial configuration.
The partition is based on the time of the automaton: for each time step $t'$, the set $X_{t'}$ consists of all variables of
the form $x_{t',r,s}$ and $y_{t',r,s}$. For each set $X_{t'}$ we require that exactly $2q-2$ variables are set to true.

The formula has the following ingredients.
\begin{itemize}
    \item \textit{At each step in time, each cell has exactly one state. Moreover, it uses a transition (unless at the first or final cell).} 
    To encode that we have at least one state, we use the expression $\bigvee_{s\in S} x_{t',r,s}$ 
    for each $t'\in [1,t-1]$ and $r\in [1,q]$. To encode that there is at least one transition, we use the clause
    $\bigvee_{z\in \cal T} y_{t',r,z}$
    for all $t'\in[1,t-1]$ and $r\in [2,q-1]$. Since we need to set exactly $2q-2$ variables to true from $X_{t'}$, and this is the total allowed amount, the pigeonhole principle shows that for each step in time $t'$ and each cell $r$, at most (and hence, exactly) one state $s$ exists for which variable $x_{t',r,s}$ is true. Similarly, for all cells apart from the first and last, there is exactly one transition $z$ for which $y_{t',r,z}$ is true.
     \item \textit{We start in the initial configuration.} We encode this using clauses with one literal $x_{0,r,s}$ whenever cell $r$ has state $s$ in the initial configuration.
    \item \textit{We end in an accepting state.} This is encoded by
    \[ \bigvee_{s\in A} \bigvee_r x_{t,r,s}. \]
    \item \textit{Left and right cells do not change.} We add clauses $x_{t',1,s_L}$ and
    $x_{t',q,s_R}$ with one literal for all time steps $t'$.
    \item \textit{If a cell has a value at a time $t'>0$, then there was a transition that caused it.} This is encoded by
    \[  x_{t',r,s} \Rightarrow \bigvee_{(s_1,s_2,s_3,s)\in {\cal T}} y_{t'-1,r, (s_1,s_2,s_3,s)}. \]
    This is expressed in conjunctive normal form as
      \[  \neg 
(x_{t',r,s})  \vee \bigvee_{(s_1,s_2,s_3,s)\in {\cal T}} y_{t'-1,r, (s_1,s_2,s_3,s)}. \]
    \item \textit{If a transition is followed, then the cell and its neighbors had the corresponding states.} For
    each time step $t'\in [1,t-1]$, cell $r\in [2,q-1]$ and transition $z=(s_1,s_2,s_3,s_4)\in {\cal T}$, we express this as
  \[ y_{t',r,z} \Rightarrow \left( x_{t',r-1,s_1} \wedge x_{t',r,s_2} \wedge x_{t',r+1,s_3}\right). 
  \]
  We can rewrite this to the three clauses $\neg (y_{t',r,z}) \vee x_{t',r-1,s_1} $, and
  $\neg (y_{t',r,z}) \vee x_{t',r,s_2} $, and $\neg (y_{t',r,z}) \vee x_{t',r+1,s_3}$. Since each `inner' cell has $y_{t,}$
\end{itemize}
The last two steps ensure that the transition chosen from the $y$-variables agrees with the states chosen from $x$-variables.

It is not hard to see that we can build the formula with  $f(k)+\log n$ space and polynomial time, and that the formula is of the required shape.
\end{proof}

A special case of the problem is when all literals that appear in the formulas $F_i$ are positive, i.e.,
we have no negations. We call this special case {\sc Chained Weighted Positive CNF-Satisfiability}.

\begin{theorem}
{\sc Chained Weighted Positive CNF-Satisfiability} is $\XNLP$-complete.
\label{theorem:chainedweightedpostivecnfsat}
\end{theorem}

\begin{proof}
We modify the proof of the previous result. Note that we can replace each negative literal by the disjunction of all other literals from a set where exactly one is true, i.e., we may replace $ \neg (x_{t',r,s})$ and $\neg (y_{t',r,z})$ by
\[  \bigvee_{s'\neq s} x_{t',r,s'} \text{ and }
 \bigvee_{z'\neq z} y_{t',r,z'} 
 \]
respectively.
The modification can be carried out in logarithmic space and polynomial time, and gives an equivalent formula. Thus, the result follows.
\end{proof}

A closer look at the proof of Theorem~\ref{theorem:chainedweightedcnfsat} shows that
$F_2 = F_3 = \cdots = F_{r-2}$, and more specifically, we have a condition on $X_1$, a condition
on $X_r$, and 
identical conditions on all pairs $X_i\cup X_{i+1}$ with $i$ from $1$ to $r-1$. Thus, we also have 
$\XNLP$-completeness of the following special case:

\medskip \begin{verse}
{\sc Regular Chained Weighted CNF-Satisfiability}\\
{\bf Input}: $r$ sets of Boolean variables $X_1, X_2, \ldots X_r$, each of size $q$; an integer $k\in \N$; Boolean formulas  $F_0$, $F_1$, $F_2$ in conjunctive normal form, where $F_0$ and $F_2$ are expressions on $q$ variables, and $F_1$ is an expression on $2q$ variables.\\
{\bf Parameter}: $k$.\\
{\bf Question}: Is it possible to satisfy the formula 
\[ F_0(X_1) \wedge \bigwedge_{1\leq i\leq r-1} F_1(X_i,X_{i+1}) \wedge F_2(X_r)\]
by setting exactly $k$ variables to true from each set $X_i$ and all others to false?
\end{verse} \medskip 

Moreover, the argument in the proof of Theorem~\ref{theorem:chainedweightedpostivecnfsat} can be applied, and thus {\sc Regular Chained Weighted Positive CNF-Satisfiability} (the variant of the problem above where all literals in $F_0$, $F_1$ and $F_2$ are positive) is $\XNLP$-complete. 

For a further simplification of our later proofs, we obtain completeness for a regular variant with only one set of constraints.

\medskip \begin{verse}
{\sc Regular Chained Weighted CNF-Satisfiability - II}\\
{\bf Given}: $r$ sets of Boolean variables $X_1, X_2, \ldots X_r$, each of size $q$; integer $k\in N$; Boolean formula $F_1$, which is in conjunctive normal form and an expression on $2q$ variables.\\
{\bf Parameter}: $k$.\\
{\bf Question}: Is it possible to satisfy the formula 
\[ \bigwedge_{1\leq i\leq r-1} F_1(X_i,X_{i+1}) \]
by setting exactly $k$ variables to true from each set $X_i$ and all others to false?
\end{verse} \medskip 

\begin{theorem}
{\sc Regular Chained Weighted CNF-Satisfiability - II} is $\XNLP$-complete.
\label{theorem:regularchainsat}
\end{theorem}

\begin{proof}
The idea of the proof is to add the constraints from $F_0$ and $F_2$ to $F_1$, but to ensure 
that they are only `verified' at the start and at the end of the chain.

To achieve this, we add variables $t_{i,j}$ for $i\in [1,r]$ and $j\in[1,r]$, with $t_{i,j}$ part of $X_i$. We increase the parameter $k$ by one. The construction is such that $t_{i,j}$ is true, if and only if $i=j$; $t_{i,1}$ implies all constraints from $F_0$, and $t_{i,r}$ implies all constraints from
$F_2$. 
The details are as follows.
\begin{itemize}
    \item We ensure that for all $i\in [1,r]$, exactly one $t_{i,j}$ is true. 
This can be done by adding a clause
\[ \bigvee_{1\leq j\leq r} t_{i,j}. \]
As the number of disjoint sets of variables that each have at least one true variable still equals $k$
(as we increased both the number of these sets and $k$ by one), we cannot have more than one true variable in the set.
    \item For all $i\in [1,r]$ and $j\in [1,r]$, we enforce the constraint $t_{i,j} \Leftrightarrow t_{i+1,j+1}$ by adding the clauses $ \neg t_{i,j} \vee t_{i+1,j+1}$ and $ t_{i,j} \vee \neg t_{i+1,j+1}$.
    \item We add a constraint that ensures that $t_{i,1}$ is false for all $i\in [2,r]$.
    This can be done by adding the clause with one literal $\neg t_{i+1,1}$ to the formula $F_1(X_i,X_{i+1})$, i.e.,
    we have a condition on a variable that is an element of the set given as second parameter.
Together with the previous set of constraints, this ensures that $t_{i,1}$ is true then $i=1$.
\item Similarly, we add a constraint that ensures that $t_{i,r}$ is false for $i<r$. This is done by
adding a clause with one literal $\neg t_{i,r}$ to $F_1(X_i,X_{i+1})$.
    \item We add a constraint of the form $t_1 \to F_0(X)$ to $F_1$; for all $i$, the variable $t_1$ is substituted by $t_{i,1}$ and $X$ by $X_i$. 
    \item We add a constraint of the form $t_r \to F_2(X)$ to $F_1$; for all $i$, the variable $t_1$ is substituted by $t_{i,1}$ and $X$ by $X_i$. 
\end{itemize}

The first four additional constraints given above ensure that for all $i \in [1,r]$
and $j\in [1,r]$, $t_{i,j}$ is true if and only if $i=j$. Thus, the fifth constraint enforces $F_0(X_1)$ (since $t_{1,1}$ has to be true); for $i>1$, this constraint has no effect.
Likewise, the sixth constraint enforces $F_2(X_r)$. Hence, the new set of constraints is equivalent to
the constraints for the first version of {\sc Regular Chained Weighted CNF-Satisfiability}.

Using standard logic operations, the constraints can be transformed to conjunctive normal form; one easily can verify the time and space bounds.
\end{proof}

Again, with a proof identical to that of Theorem~\ref{theorem:chainedweightedpostivecnfsat}, we can show that the variant with only positive literals ({\sc Regular Chained Weighted Positive CNF-Satisfiability - II}) is $\XNLP$-complete.

From the proofs above, we note that each set of variables $X_i$ can be partitioned into $k$ subsets, and a solution has exactly one true variable for each subset, e.g., for each $t'$, exactly one
$x_{t',r,s}$ is true in the proof of Theorem \ref{theorem:chainedweightedcnfsat}. This still holds after the modification in the proofs of the later results. We define the following variant.

\medskip \begin{verse}
{\sc Partitioned Regular Chained Weighted CNF-Satisfiability}\\
{\bf Input}: $r$ sets of Boolean variables $X_1, X_2, \ldots X_r$, each of size $q$; an integer $k\in \N$; Boolean formula $F_1$, which is in conjunctive normal form and an expression on $2q$ variables; for each $i$,
a partition of $X_i$ into $X_{i,1}, \ldots, X_{i,k}$ with for all $i_1,i_2$, $j$: $|X_{i_1,j}| = |X_{i_2,j}|$.\\
{\bf Parameter}: $k$.\\
{\bf Question}: Is it possible to satisfy the formula 
\[ \bigwedge_{1\leq i\leq r-1} F_1(X_i,X_{i+1}) \]
by setting from each set $X_{i,j}$ exactly $1$ variable to true and all others to false?
\end{verse} \medskip 

We call the variant with only positive literals {\sc Partitioned Regular Chained Weighted Positive CNF-Satisfiability}.

\begin{corollary}
{\sc Partitioned Regular Chained Weighted CNF-Satis\-fia\-bi\-li\-ty} and {\sc Partitioned Regular Chained Weighted Positive CNF-Satis\-fiability} are
$\XNLP$-complete.
\end{corollary}

%%%%%%%%%% Chained Multicolored Clique
\subsection{Chained Multicolored Clique}
\label{subsection:chainedmcclique}
The {\sc Multicolored Clique} problem is an important tool to prove fixed parameter intractability
of various parameterized problems. It was independently introduced by Pietrzak~\cite{Pietrzak03} (under
the name {\sc Partitioned Clique}) and by Fellows et al.~\cite{FellowsHRV09}. 

In this paper, we introduce a chained variant of {\sc Multicolored Clique}. In this variant, we ask to find a sequence of cliques, that are overlapping with the previous and next clique in the chain.

\medskip \begin{verse}
  {\sc Chained Multicolored Clique}\\
  {\bf Input:} Graph $G=(V,E)$; partition of $V$ into sets $V_1, \ldots, V_r$, such that
  for each edge $\{v,w\}\in E$, if $v\in V_i$ and $w\in V_j$, then $|i-j|\leq 1$; 
  function $f: V \rightarrow \{1,2,\ldots, k\}$.\\
  {\bf Parameter:} $k$.\\
  {\bf Question:} Is there a subset $W\subseteq V$ such that for each $i\in [1,r]$,
  $W \cap (V_i \cup V_{i+1})$ is a clique, and for each $i\in [1,r]$ and each $j\in [1,k]$,
  there is a vertex $w\in V_i\cap W$ with $f(w)=j$?
\end{verse} \medskip
Thus, we have a clique with $2k$ vertices in $V_i\cup V_{i+1}$ for each $i\in [1,r-1]$, with for each color a vertex with that color in $V_i$ and a vertex with that color in $V_{i+1}$. Importantly, the same vertices in $V_i$ are chosen in the clique for $V_{i-1}\cup V_i$ as for
$V_i \cup V_{i+1}$ for each $i\in [2,r-1]$. Below, we call such a set a {\em chained multicolored clique}.

\begin{theorem}
{\sc Chained Multicolored Clique} is $\XNLP$-complete.
\label{theorem:chainedmulticoloredclique}
\end{theorem}
\begin{proof}
Membership in $\XNLP$ is easy to see: iteratively guess for each $V_i$ which vertices belong to the clique. We only need to keep the clique vertices in $V_{i-1}$ and $V_i$ in memory.

We now prove hardness via a transformation from {\sc Partitioned Regular Chained Weighted Positive CNF-Satisfiability}.
We are given:
\begin{itemize}
    \item $r$ sets of Boolean variables $X_1, X_2, \ldots X_r$, each of size $q$;
    \item an integer $k\in \N$; for each $i$, a partition of $X_i$ into $X_{i,1}, \ldots, X_{i,k}$ with for all $i_1,i_2$, $j$: $|X_{i_1,j}| = |X_{i_2,j}|$;
    \item Boolean formula $F_1$, which is in conjunctive normal form using only positive literals and an expression on $2q$ variables.
\end{itemize}
We need to decide if it is possible to satisfy the formula 
\[ \bigwedge_{1\leq i\leq r-1} F_1(X_i,X_{i+1}) \]
by setting from each set $X_{i,j}$ exactly $1$ variable to true and all others to false.

We build an equivalent instance of {\sc Chained Multicolored Clique}. 
We create a graph with the following vertices. For each $i\in [r-1]$ and for each clause $c$ in $F_1$, we create a vertex set $V_{i,c}$. Informally, this serves to check whether the clause $c$ is satisfied by $X_i\cup X_{i+1}$, so we need to `track' which of those are set to true. For each $j\in [k]$, we add the following vertices to $V_{i,c}$. 
\begin{itemize}
\item We add two vertices $a_{c,i,j}$ and $b_{c,i,j}$. We give  these the color $2k+1$. Informally, choosing the vertex $a_{c,i,j}$ corresponds to satisfying clause $c$ using a vertex from $X_{i,j}$; similarly, choosing the vertex $b_{c,i,j}$ corresponds to satisfying clause $c$ using a vertex from $X_{i+1,j}$. We refer to these as \emph{clause checking} vertices.
    \item For each $x
\in X_{i,j}$, we add a vertex $v_{x,i,j,c}$ which we give color $j$. These vertices keep track of the variable $x$  from $X_{i,j}$ which is assigned True.  We refer to these as \emph{current selection} vertices and say $v_{x,i,j,c}$ \emph{selects} $x$.
    \item For each $y
\in X_{i+1,j}$, we add a vertex $w_{y,i,j,c}$ which we give color $j+k$. These vertices keep track of the variable $y$  from $X_{i+1,j}$ which is assigned True. We refer to these as \emph{next selection} vertices and say $w_{y,i,j,c}$ \emph{selects} $y$.
\end{itemize}
This defines the vertex set and vertex coloring (with $2k+1$ colors, the new parameter).

We place an arbitrary order on the clauses $c_1,\dots,c_f$ in $F_1$ and the partition of the vertex set gets the order
\[
(V_{1,c_1},V_{1,c_2},\dots,V_{1,c_f},V_{2,c_1},V_{2,c_2}\dots,\dots,V_{r-1,c_f}).
\]
Next, we define the edges, which will only be between vertices in the same or consecutive parts (according to the partition order above). The idea is that vertices that model different selection choices have no restrictions on each other; however, if we choose $v_{x,i,j,c}$ then we should also choose  $v_{x,i,j,c'}$ for all future $c'$, for example. 

We first describe the adjacencies between the (current or next) selection vertices. 
\begin{itemize}
    \item \emph{Case 1: vertices with the same $i\in [r-1]$.}
    If selection vertices $u
\in V_{i,c}$ and $v\in V_{i,c'}$ are placed in the same part ($c=c'$), or consecutive parts, then they are adjacent if and only if 
\begin{itemize}
    \item they have different colors (from $[2k]$), or
    \item they select the same vertex (i.e. $v=v_{x,i,j,c}$ and $u=v_{x,i,j,c'}$ for some $x$, or $u=w_{y,i,j,c}$ and $v=w_{y,i,j,c'}$).
\end{itemize}
\item \emph{Case 2: one vertex has $i\in [r-2]$ and the other $i+1$.}
The vertices $u
\in V_{i,c_r}$ and $v\in V_{i+1,c_1}$ are adjacent if 
\begin{itemize}
    \item $u$ is a current selection vertex (`v-type'), or
    \item $v$ is a next selection vertex  (`w-type'), or
    \item $u=w_{x,i,j,c_r}$ and $v=v_{x',i,j',c_1}$ where $j\neq j'$ or $x=x'$.
\end{itemize}
\end{itemize}
Note that, as we wanted, if vertices are in a consecutive part, and both model selection from the same set $X_{i,j}$, then they are only adjacent if they select the same vertex.

Next, we define adjacencies to $a_{c,i,j}$ and $b_{c,i,j}$ in order to check whether $c$ is satisfied. For $i\in [r-1],j\in [k]$ and $c$ a clause in $F_1$,
\begin{itemize}
    \item the vertices $a_{c,i,j}$ and $b_{c,i,j}$ are adjacent to all vertices from the previous and next part;
    \item the vertex $a_{c,i,j}$ is adjacent to $v_{x,i,j,c}\in V_i$ if and only if $x\in X_{i,j}$ appears in $c$ (as a positive literal), and it is adjacent to all vertices of the form $w_{y,i,j,c}$;
    \item the vertex $b_{c,i,j}$ is adjacent to $w_{y,i,j,c}\in V_{i+1}$ if and only if $y\in X_{i+1,j}$ appears in $c$ (as a positive literal), and it is adjacent to all vertices of the form $v_{x,i,j,c}$.
\end{itemize}
Note that there are only edges between vertices in consecutive parts and that with $f$ is the total number of clauses and $\ell=\max_{i,j}|X_{i,j}|$, our total number of created vertices is at most $rkf(\ell+2)$, for $f$ the number of clause in $F_1$.

We claim the created instance admits a multicolored clique if and only if the original instance was satisfiable.  

Suppose first that we are given a satisfying assignment, given by a choice of $t_{i,j}\in X_{i,j}$ for each $i\in [r]$ and $j\in [k]$ (i.e. the variables to be set to True). For $i\in [r], j\in [k]$ and clause $c$ from $F_1$, we select $v_{t_{i,j},i,j,c}$ if $i\leq r-1$ and we also select $w_{t_{i,j},i-1,j,c}$ if $i\geq 2$. These vertices are adjacent (for any choice of $t_{i,j}$) so this is a multicolored clique except for missing the color $2k+1$ in all parts. 
For each $i\in [r-1]$ and clause $c$, there is a choice of $j\in [k]$ such that either $t_{i,j}$ or $t_{i+1,j}$ appears in $c$, since $c$ is satisfied. We choose $a_{c,i,j}$ or $b_{c,i,j}$ respectively. This gives the desired multicolored clique (with all $2k+1$ colors now). 

Conversely, if we are given a multicolored clique, then for $i\in [r]$ and $j\in [k]$, let $v_{x,i,j,c_1}$ be the chosen vertex of color $j$ from $V_{i,c_1}$. We set $x\in X_{i,j}$ to True. This gives an assignment which sets the right number of variables to True and we need to check if it satisfies all the clauses.
Let $i\in [r-1]$ and $c$ a clause in $F_1$.
There has to be a vertex $u$ of color $2k+1$ in $V_{i,c}$, so $u$ is a clause checking vertex.
\begin{itemize}
    \item Case 1: $u$ is of the form $a_{c,i,j}$. It is adjacent to the chosen vertex of color $j$ from $V_{i,c}$, of the form $v_{x,i,j,c}$ (since we have a clique). By definition of the edges, this means $x\in X_{i,j}$ appears in $c$ and so we just need to argue that $x$ has been set to True. This follows from an inductional argument: it is immediate if $c=c_1$ and if $c=c_b$ for $b>1$, then the only vertex in the previous part of color $j$ that is adjacent to $v_{x,i,j,c}$ is $v_{x,i,j,c_{b-1}}$. So $v_{x,i,j,c_{b-1}}$ must be part of the clique, and continuing, $v_{x,i,j,c_{1}}$ must be part of the clique and indeed we set $x$ to True.
    \item  Case 2: $u$ is of the form $b_{c,i,j}$. This must be adjacent to a chosen vertex of the form $w_{x,i,j,c}$ of color $j+k$. Again, the clause will be satisfied if we set $x\in X_{i+1,j}$ to True, which we did if we chose the vertex $v_{x,i-1,j,c_1}$. This follows by a similar inductional argument.
\end{itemize}
All clauses are satisfied so indeed there is a satisfying assignment. 

This shows the required equivalence and proves the claim. Finally, we note the reduction is easily performed in $f(k)+\log(rf\ell)$ space.
\end{proof}

A simple variation is the following. We are given a graph $G=(V,E)$, a partition of $V$ into sets $V_1, \ldots, V_r$ with the property that for each edge $\{v,w\}\in E$, if $v\in V_i$ and $w\in V_j$ then $|i-j|\leq 1$, and a coloring function $f: V \rightarrow \{1, 2, \ldots, k\}$. A {\em chained multicolored independent set} is an independent set $S$ with the property that for each $i\in [1,r]$
and each color $j\in [1,k]$, the set $S$ contains exactly one vertex $v\in V_i$ of color $f(v)=j$.
The {\sc Chained Multicolored Independent Set} problem asks for the existence of such a chained multicolored independent set, with the number of colors $k$ as parameter. We have the following simple corollary.

\begin{corollary}
{\sc Chained Multicolored Independent Set} is $\XNLP$-com\-plete.
\end{corollary}
\begin{proof}
This follows directly from Theorem~\ref{theorem:chainedmulticoloredclique}, by observing that
the following `partial complement' of a partitioned graph $G=(V_1 \cup \cdots V_r, E)$ can be constructed in logarithmic space: create an edge $\{v,w\}$ if and only if
there is an $i$ with $v\in V_i$ and $w\in V_i \cup V_{i+1}$, $v\neq w$ and $\{v,w\}\not\in E$.
\end{proof}

%%%%%%%%%%%%%%%
%% BINARY CSP
%%%%%%%%%%%%%
\subsection{Binary CSP parameterized by pathwidth}
\label{subsection:binarycsp}
An instance of \textsc{Binary CSP} is a triple
$$I=(G,\{D(u)\colon u\in V(G)\},\{C(u,v)\colon uv\in E(G)\}),$$
where
\begin{itemize}
    \item $G$ is an undirected graph, called the {\em{Gaifman graph}} of the instance;
    \item for each $u\in V(G)$, $D(u)$ is a finite set called the {\em{domain}} of $u$; and
    \item for each $uv\in E(G)$, $C(u,v)\subseteq D(u)\times D(v)$ is a binary relation called the {\em{constraint}} at~$uv$. Throughout this paper, we apply the convention that $C(v,u)=\{(b,a)\,\mid\,(a,b)\in C(u,v)\}$.
\end{itemize}
A {\em{satisfying assignment}} for an instance $I$ is a function $\eta$ that maps every variable $u$ to a value $\eta(u)\in D(u)$ such that for every edge $uv$ of $G$, we have $(\eta(u),\eta(v))\in C(u,v)$. The \textsc{Binary CSP} problem asks, for a given instance $I$, whether $I$ is {\em{satisfiable}}, that is, there is a satisfying assignment for $I$.

For \textsc{Binary CSP} parameterized by pathwidth, we assume a path decomposition of width $k$ of the Gaifman graph $G$ is also given as part of the input and consider $k$ as our parameter.
\begin{theorem}
\label{thm:BinCSP}
    \textsc{Binary CSP} is XNLP-complete when parameterized by the following parameters:
    \begin{enumerate}
        \item pathwidth;
        \item pathwidth + maximum degree;
        \item cutwidth;
        \item bandwidth.
    \end{enumerate}
\end{theorem}
\begin{proof}
We first prove membership when parameterized by pathwidth, using a `standard' dynamic programming approach.
If the given path decomposition is $(X_1,X_2,\dots,X_\ell)$, then for $i=1,\dots,n$, 
\begin{itemize}
    \item we (non-deterministically) guess an element of $f_i(v)\in D(v)$ for each $v\in X_i$ (which we store in memory),
    \item we check constraints: check if $(f_i(u),f_i(v))\in C(u,v)$ for all $u,v\in X_i$ with $uv\in E(G)$ (else fail),
    \item we check consistency: if $v\in X_{i-1}$, we check whether $f_i(v)=f_{i-1}(v)$ (else fail),
    \item we free up memory (forget $f_{i-1}(v)$ for all $v\in X_{i-1}$).
\end{itemize} 
The total memory used is $O(k \log n)$ bits, with $k$ the pathwidth and $n$ the number of vertices. The described non-deterministic algorithm also runs in fpt time. Membership for the other parameters follows analogously.

Next, we prove XNLP-hardness via a transformation from {\sc Chained Multicolored Clique}.

Let $G=(V,E)$ be a $k$-colored graph with partition $V_1, \ldots, V_r$ of the vertex set, such that edges are only between the same or consecutive parts. 

For each $i\in [r]$ and $j\in [k]$, we create a vertex $u_{i,j}$. Let $U_i=\{u_{i,j}:j\in [k]\}$.
A vertex $u\in U_i$ is adjacent to $v\in U_{i'}$ if and only if $|i-i'|\leq 1$. This defines the Gaifman graph.
So 
\[
(X_1,\dots,X_{r-1}):=(U_1\cup U_2, U_2\cup U_3, \dots, U_{r-1}\cup U_r)
\]
gives a path decomposition of the Gaifman graph of width at most $2k$. Moreover, the maximum degree of the graph is at most $3k$ and the total number of vertices is $rk$.

For $i\in [r]$ and $j\in [k]$, the domain $D(u_{i,j})$ is given by the set of vertices of color $j$ in $V_i$.
Adjacent vertices $u_{i,j}$ and $u_{i',j'}$ have the following constraint set:
\[
C(u_{i,j},u_{i',j'})=\{(v,v')\in D(u_{i,j} )\times D(u_{i',j'} ): vv'\in E(G)\}.
\]
In words, they may only be assigned to adjacent vertices (from $V_i$ of color $j$ and $V_{i'}$ of color $j'$ respectively, by definition of the domains).

Once the construction is understood, the equivalence is direct: any satisfying assignment is a multicolored clique and vice versa. 

The Gaifman graph in our reduction has 
bandwidth at most $2k-1$: take the linear
ordering $f$ with $f(u_{i,j})= (i-1)\cdot k+j$.
(I.e., we first number the vertices in $U_1$, then in $U_2$, etc. As each vertex is only adjacent to the vertices in its own set $U_i$,
and the previous and next sets $U_{i-1}$, and
$U_{i+1}$, the bandwidth of this ordering is
$2k-1$.) This linear ordering $f$ also has
cutwidth $O(k^2)$, and the hardness for all
claimed parameters now follows.
\end{proof}
We remark that we will later show that \textsc{List Coloring} is also XNLP-complete parameterized by pathwidth. The simple reduction however blows up the maximum degree, which is to be expected because the problem becomes fpt when both the pathwidth and the maximum degree are bounded. This means that we cannot expect to extend the result above to \textsc{List Coloring} for the other three parameters.

%%%%%%%%%%% The Non-decreasing Non-deterministic Checking Counter Machines %%%%%%%%
\subsection{Non-decreasing counter machines}
\label{subsection:nncc}
In this subsection, we introduce a new simple machine model, which can also capture the computational power of
$\XNLP$ (see Theorem~\ref{theorem:nnccm}). This model will be a useful stepping stone when proving $\XNLP$-hardness reductions in Section~\ref{section:applications}.

A {\em Nondeterministic Nondecreasing Checking Counter Machine} (or: NNCCM) is described by a
3-tuple $(k, n, s)$, with $k$ and $n$ positive integers, and $s = (s_1, \ldots, s_r)$ a sequence of 4-tuples (called {\em checks}). For each $i\in\{1,\ldots,r\}$, the 4-tuple $s_i$ is of the form
$(c_1,c_2, r_1, r_2)$ with $c_1, c_2\in \{1,2,\ldots, k\}$ positive integers and $r_1,r_2 \in \{0, 1, 2, \ldots, n \}$ non-negative integers. These model the indices of the counters and their values respectively.

An NNCCM  $(k, n, s)$ with $s = (s_1, \ldots, s_r)$ works as follows. The machine has $k$ counters that
are initially $0$. For $i$ from $1$ to $r$, the machine first sets each of the counters to any integer that
is at least its current value and at most $n$. After this, the machine performs the $i$th check $s_i = (c_1,c_2,r_1, r_2)$: if the value of the $c_1$th counter equals $r_1$ and the value of
the $c_2$th counter equals $r_2$, then we say the $i$th check rejects and the machine halts and rejects. When the machine has not rejected
after all $r$ checks, the machine accepts.

For example, we may have $n=2$ and $k=3$, with checks \[
s=((1,2,0,1),(2,3,1,1),(1,3,0,2)).
\]
We pass $s_1$ since $(c_1,c_2)=(0,0)\neq (0,1)$. We then increase $c_1$ to $1$ and increase $c_3$ to $2$. We succeed $s_2$ and $s_3$ since $(c_2,c_3)=(0,2)\neq (1,1)$ and $(c_1,c_3)=(1,2)\neq (0,2)$.  The machine accepts since none of the checks rejected. (Many variations on this also work.) An example which always rejects is $n=1,k=2$ and 
\[
s=((1,2,0,0),(1,2,1,0),(1,2,0,1),(1,2,1,1)).
\]

The nondeterministic steps can be also described as follows. Denote the value of the $c$th counter 
when the $i$th check is done by $c(i)$. We define $c(0)=0$. For each $i\in \{1,\dots,r\}$,
$c(i)$ is an integer that is nondeterministically chosen from $[c(i-1), n]$. 

We consider the following computational problem.
\medskip \begin{verse}
{\sc Accepting NNCCM}\\
{\bf Given:} An NNCCM $(k, n, s)$ with all integers given in unary.\\
{\bf Parameter:} The number of counters $k$.\\
{\bf Question:} Does the machine accept?
\end{verse} \medskip

\begin{theorem}
{\sc Accepting NNCCM} is $\XNLP$-complete.
\label{theorem:nnccm}
\end{theorem}
\begin{proof}
We first argue that {\sc Accepting NNCCM} belongs to $\XNLP$. We simulate the execution of the
machine. At any point, we store $k$ integers from $[0,n]$ that give the current values of our counters, as well as the index $i\in [1,r]$ of the check that we are performing. This takes only $O(k\log n+\log r)$ bits. When we perform the check $s_i=(c_1,c_2,r_1,r_2)$, we store the values $c_1,c_2,r_1,r_2$ in order to perform the check using a further $O(\log n)$ bits. So we can simulate the machine using $O(k\log n+\log r)$ space. The running time is upper bounded by some function of the form $f(k)\cdot \poly(n,r)$.

We now prove hardness via a transformation from {\sc Chained Multicolored Clique}.
We are given a $k$-colored graph with vertex sets $V_1, \ldots, V_s$.
By adding isolated vertices if needed, we may assume that, for each $i\in[1,s]$, the set $V_i$ contains exactly $m$ vertices of each color. We will assume that $\ell$ is even; the proof is very similar for odd $\ell$.
We set $n = ms$. 
 
We create $4k$ counters: for each colour $i\in [1,k]$, there are counters $c_{i,1,+}$, $c_{i,1,-}$, $c_{i,0,+}$, $c_{i,0,-}$. We use the counters $c_{i,1,\pm}$ for selecting vertices from sets $V_j$ with $j$ odd and the counters $c_{i,0,\pm}$ for selecting vertices from sets $V_j$ with $j$ even.

The intuition is the following. 
We increase the counters in stages, where in stage $j$ we model the selection of the vertices from $V_j$. Say $j$ is even. We increase the counters $c_{i,0,+}$, $c_{i,0,-}$ to values within $[jm+1, (j+1)m]$. Since counters may only move up, there can be at most one $\ell\in [1,m]$ for which the counters at some point take the values $c_{i,0,+}=jm+\ell$ and $c_{i,0,-}=(j+1)m+1-\ell$. We enforce that such an $\ell$ exists and interpret this as placing the $\ell$th vertex of color $i$ in $V_j$ into the chained multicolored clique.

We use the short-cut $(c_1,c_2,R_1,R_2)$ for the sequence of checks $((c_1,c_2,r_1,r_2):r_1\in R_1,~r_2\in R_2)$ performed in lexicographical order, e.g. for $R_1=\{1,2\}$ and $R_2=\{1,2,3\}$, the order is $(1,1),(1,2),(1,3),(2,1),(2,2),(2,3).$

For each $j\in [1,s]$, the $j$th \textit{vertex selection check} confirms that for each $i \in [1,k]$, $(c_{i,par,+},c_{i,par,-})$ is of the form $(jm+\ell,(j+1)m+1-\ell)$ for some $\ell\in [1,m]$, where $par$ denotes the parity of $j$. For each $i\in [1,k]$, we set $c_1=c_{i,par,+}$ and $c_2=c_{i,par,-}$, and perform the following checks in order.
\begin{enumerate}
    \item $(c_1,c_2,[0,jm-1],[0,n])$;
    \item $(c_1,c_2,[0,n],[0,jm-1])$;
    \item for $\ell\in [1,m]$, $(c_1, c_2, \{jm+\ell\}, [jm,(j+1)m] \setminus \{(j+1)m+1 - \ell \})$;
    \item $(c_1,c_2,[(j+1)m+1,n],[0,n])$;
    \item $(c_1,c_2,[0,n],[(j+1)m+1,n])$.
\end{enumerate}
Suppose all the checks succeed. After the second check, $c_1$ and $c_2$ are both at least $jm$, and before the last two checks, $c_1$ and $c_2$ are both at most $(j+1)m$. The middle set of checks ensure that there is some $\ell$ for which the $c_1$th counter and $c_2$th counter have been simultaneously at the values $jm+\ell$ and $(j+1)m+1-\ell$ respectively. We say the check \emph{chooses} $\ell$ for $c_1$ and $c_2$. Since the counters can only move up, this $\ell$ is unique.

This is used as a subroutine below, where we create a collection of checks such that the corresponding NNCCM accepts if and only if the $k$-colored graph has a chained multicolored clique.
For $j=1$ to $s$, we do the following.
    \begin{itemize}
        \item Let $par \equiv j \mod 2$ denote the parity of $j$ and
          let $par'\in \{0,1\}$ denote the opposite parity.
        \item We perform a $j$th vertex selection check.
        \item We verify that all selected vertices in $V_j$ are adjacent. Let $uu'$ be a non-edge with $u,u'\in V_j$. Let $\ell,i,\ell',i'$ with $i\neq i'$ be such that $u$ is the $\ell$th vertex of color $i$ in $V_j$ and $u'$ is the $\ell'$th vertex of color $i'$. We add the check $(c_{i,par,+},c_{i',par,+},jm+\ell,jm+\ell')$. This ensures that we do not put both $u$ and $u'$ in the clique of $V_j$.
        \item If $j>1$, then we verify that all selected vertices in $V_j$ are adjacent to all selected vertices
        in $V_{j-1}$. 
        Let $uu'$ be a non-edge with $u\in V_{j}$ and $u'\in V_{j-1}$ and let $\ell,i,\ell',i'$ be as above. We add the check $(c_{i,par,+},c_{i',par',+},jm+\ell,jm+\ell')$ to ensure that we do not put both $u$ and $u'$ into the clique. 
        \item We finish with another $j$th vertex selection check. If $j>1$, we  also do a $(j-1)$th vertex selection check. This ensures our counters are still `selecting vertices' from $V_{j-1}$ and $V_j$.
    \end{itemize}
We now argue that the set of checks created above accepts if and only if the graph contains a chained multicolored clique. Suppose first that such a set $W\subseteq V$ exists for which $W\cap (V_j\cup V_{j+1})$ forms a clique for all $j$ and $W$ contains at least one vertex from $V_j$ of each color. We may assume that $W_j=W\cap V_j$ is of size $k$ for each $j$. Let $j\in [1,s]$ be given of parity $par$ and for each $i\in [k]$, let the $f(i)$th vertex of $V_j$ of color $i$ be in $W_j$. Before the first $j$th vertex selection check, we move the counters $(c_{i,par,+},c_{i,par,-})$ to $(jm+f(i),(j+1)m-1-f(i))$, and these will be left there until the first $(j+2)$th vertex selection check. This will ensure that all the checks accept.

Suppose now that all the checks accept. We first make an important observation. Let two counters $c_1$ and $c_2$ be given. If in an iteration above, the first vertex selection check chooses $\ell$ for $c_1$ and $c_2$ and the second vertex selection check chooses $\ell'$, then it must be the case that $\ell=\ell'$. Indeed, we cannot increase $c_1$ or $c_2$ beyond $(j+1)m$ before the last vertex selection check, and they need to be above $jm$ due to the first. If the first vertex selection check selects $\ell$, then the $c_1$the counter is at least $jm+\ell$, so in order for it to be $jm+\ell'$ in the second check, we must have $\ell'\geq \ell$. Considering the value of the $c_2$th counter, we also find $\ell\geq \ell'$ and hence $\ell=\ell'$. In particular, the counters cannot have moved between the two vertex selection checks.

It is hence well-defined to, for $j\in [1,s]$ of parity $par$, let $W_j$ be the set of vertices that are for some color $i\in [1,k]$ the $\ell$th vertex of color $i$ in $V_j$, for $\ell$ the unique value that is selected by a $j$th vertex selection check for $c_{i,par,+}$ and $c_{i,par,-}$. We claim that $W=\cup_{j=1}^s W_j$ is our desired multicolored chained clique. It contains exactly one vertex per color from $V_j$. Suppose $u\in W_j$ and $u'\in W_{j-1}\cup W_j$ are not adjacent and distinct. Let $i,par,\ell$ be such that $u$ is the $\ell$th color of parity $par$ in $V_j$ for $j$ of parity $par$, and similar for $i',par',\ell'$. Then at the $j$th iteration, the check $(c_{i,par,+},c_{i',par,+},jm+\ell,jm+\ell')$ has been performed (because $uu'$ is a non-edge). Since $u,u'\in W$ and this check is done between vertex selection checks, the counters have to be at those values. This shows that there is a check that rejects, a contradiction. So $W$ must be a chained multicolored clique. 
\end{proof}

The {\sc Accepting NNCCM} problem appears to be a very useful tool for giving $\XNLP$-hardness proofs. 
Note that one step where the $k$ counters can be increased to values at most $n$ can be replaced by
$kn$ steps where counters can be increased by one, or possibly a larger number, again to at most $n$.
This modification is used in 
some of the proofs in the following section.

\section{Applications}
\label{section:applications}

In this section, we consider several problem, which we prove to be $\XNLP$-complete. Using
the building blocks of the previous section,
we provide $\XNLP$-completeness results for
the following problems:

\begin{itemize}
    \item Subsection~\ref{subsection:listcoloringpathwidth}: {\sc List Coloring} and \textsc{Precoloring Extension} parameterized by pathwidth.
    \item Subsection~\ref{subsection:logpathwidth}: {\sc Dominating Set} and \textsc{Independent Set} parameterized by logarithmic pathwidth.
    \item Subsection~\ref{subsection:scheduling}: {\sc Scheduling with Precedence Constraints} parameterized by the number of machines and the partial order width.
    \item Subsection~\ref{subsection:uepath}: {\sc Uniform Emulations of Weighted Paths}.
    \item Subsection~\ref{subsection:bandwidth}: {\sc Bandwidth}. 
    \item Subsection~\ref{section:fsaintersection}: %\todo{CG: 4.6 is also there!}:
    {\sc Acyclic Finite State Automata Intersection}.
\end{itemize}

%%%%%%%%% List coloring parameterized by pathwidth %%%%%%%%%%%%%

\subsection{List Coloring Parameterized by Pathwidth}
\label{subsection:listcoloringpathwidth}
In this section, we show that \textsc{List Coloring} and \textsc{Precoloring Extension} are XNLP-complete when parameterized by pathwidth.

In the \textsc{List Coloring} problem,
we are given a graph $G=(V,E)$, a finite set of colors $\mathcal{C}$,
and for each vertex $v\in V$, a subset of the colors $C(v)\subseteq \mathcal{C}$, and ask if there is a function $f: V\rightarrow \mathcal{C}$, such that each vertex gets assigned a color from its own set: $\forall v\in V: f(v)\in C(v)$, and the endpoints of each edge have different colors: $\forall vw\in E: f(v)\neq f(w)$. We call such a function $f$ a \textit{list coloring} for $G$.

In the \textsc{Precoloring Extension} problem, 
we are given a graph $G=(V,E)$, a set of colors $\mathcal{C}$, a
set of precolored vertices $W\subseteq V$, a function $p: W \rightarrow \mathcal{C}$, and ask for a coloring $f: V \rightarrow \mathcal{C}$ that
extends $p$ (for all $w\in W: f(w)=p(w)$), such that the endpoints of 
each edge have different colors: $\forall vw\in E: f(v)\neq f(w)$.

The {\sc List Coloring} and {\sc Precoloring Extension} problems
parameterized by tree\-width or pathwidth belong to XP \cite{JansenS97}. 
Fellows et al.~\cite{FellowsFLRSST11} have shown that {\sc List Coloring} and {\sc Precoloring Extension}
parameterized by treewidth are $\mathrm{W}[1]$-hard. (The transformation in their paper also works for parameterization by pathwidth.) 
We strengthen this result by showing that {\sc List Coloring} and {\sc Precoloring Extension} parameterized by pathwidth are $\XNLP$-complete. Note that this
result also implies $\mathrm{W}[t]$-hardness of the problems for all integers $t$.
\begin{theorem}
{\sc List Coloring} parameterized by pathwidth is $\XNLP$-complete.
\label{theorem:listcoloring}
\end{theorem}
\begin{proof}
Membership follows analogously to \textsc{Binary CSP} (or by using that {\sc List Coloring}  is a special case of this).

In order to prove hardness, we reduce from \textsc{Binary CSP}. Let an instance $(G,(D(u))_{u\in V(G)},(C(u,v))_{uv\in E(G)})$ of \textsc{Binary CSP} be given. We create a copy $G'$ of $G$ and also copy the domains: if $v'\in V(G')$ is a copy of $v\in V(G)$, then $L(v')=D(v)$. 

For an edge $uv\in E(G)$, $u',v'\in V(G')$ denote the copies of $u,v\in V(G)$. For each pair $(c,d)\in D(u)\times D(v)\setminus C(u,v)$, we add a \emph{helper} vertex $h_{uv,c,d}$ to $V(G')$ with list $\{c,d\}$ and make it adjacent to $u'$ and $v'$.

A list coloring $\alpha$ of the copy of $G$ within $G'$ is exactly an assignment of the original instance (which satisfies the domains). It is easily checked that such a list coloring extends to the helper vertices of $G'$ if and only if $(\alpha(u'),\alpha(v'))\in C(u,v)$ for all edges $uv$ in $G$. This means there exists a list coloring if and only if there is a satisfying assignment for the original instance.
\end{proof}

We deduce the following result from a well-known transformation from {\sc List Coloring}.

\begin{corollary}
{\sc Precoloring Extension} parameterized by pathwidth is \linebreak $\XNLP$-complete.
\label{theorem:precoloring}
\end{corollary}

\begin{proof}
Membership follows immediately since {\sc Precoloring Extension} can be seen as a special case  of {\sc List Coloring}  in which the precolored vertices have lists of size one.

We prove hardness via a reduction from {\sc List Coloring}.
Consider an instance of {\sc List Coloring} with graph $G=(V,E)$ and color lists $C(v)$
for all $v\in V$. Let ${\cal C} = \bigcup_{v\in V} C(v)$ be the set of all colors.
For each vertex $v$ with color list $C(v)$, for each color $\gamma \in {\cal C}\setminus C(v)$, add a new vertex that is only adjacent to $v$ and is
precolored with $\gamma$. The original vertices are not precolored. 
It is easy to see that we can extend the precoloring of the resulting graph to a proper coloring, if and only if the instance of {\sc List Coloring} has
a solution.

The procedure increases the pathwidth by at most one. Indeed, if we are given
a path decomposition of $G$, then we can build the path decomposition of the
resulting graph as follows. We iteratively visit all the bags $X_i$ of the path decomposition. Let $v\in X_i$ be a vertex for which $X_i$ is the first bag it appears in (so $i=1$ or $v\not\in X_{i-1}$). We create a new bag $X_i \cup \{w\}$ for each new neighbor $w$ of $v$ and place this bag somewhere in between $X_i$ and $X_{i+1}$. 
This transformation can be easily executed with $k+O(\log n)$ bits of memory.
\end{proof}

From Theorem~\ref{theorem:listcoloring}, and Corollary~\ref{theorem:precoloring} we can also directly conclude that {\sc List Coloring} and {\sc Precoloring Extension} are $\XNLP$-hard when parameterized by the treewidth. 
However, we expect that these problems are not members of XNLP; they are shown to be complete for the class XALP in \cite{BodlaenderGJPP22}, and we conjecture that XNLP is a proper subset
of XALP.

%%%%%%%%%% Problems with logarithmic pathwidth

\subsection{Logarithmic Pathwidth}
\label{subsection:logpathwidth}
There are several well known problems that can be solved in time $O(c^k n)$ for a constant $c$ on graphs of pathwidth or treewidth at most $k$.
Classic examples are {\sc Independent Set} and {\sc Dominating Set} (see e.g., \cite[Chapter 7.3]{CyganFKLMPPS15}), but there are many others, 
e.g., \cite{BodlaenderCKN15,TelleP97}.

In this subsection, rather than bounding the pathwidth by a constant, we allow the pathwidth to be linear in the logarithm of the number of vertices of the graph. We consider the following problem.
\medskip \begin{verse}
{\sc Log-Pathwidth Dominating Set}\\
{\bf Input:} Graph $G=(V,E)$, path decomposition of $G$ of width $\ell$, integer $K$. \\
{\bf Parameter:} $\left\lceil \ell / \log |V| \right\rceil$. \\
{\bf Question:} Does $G$ have a dominating set of size at most $K$?
\end{verse} \medskip
The independent set variant and clique variants are defined analogously.
\begin{theorem}
{\sc Log-Pathwidth Dominating Set} is $\XNLP$-complete.
\end{theorem}
\begin{proof}
It is easy to see that the problem is in $\XNLP$ --- run the standard dynamic programming algorithm for {\sc Dominating Set} on graphs with bounded pathwidth, but instead of computing full tables, guess the table entry for each bag.

Hardness for $\XNLP$ is shown with help of a reduction from {\sc Partitioned Regular Chained Weighted Positive CNF-Satisfiability}.

Suppose we are given $F_1$, $X_1$, \ldots, $X_{r}$, $q$, $k$, $X_{i,j}$ ($1\leq i\leq r$, $1\leq j\leq k$) as instance of the {\sc Partitioned Regular Chained Weighted Positive CNF-Satisfiability} problem.

We assume that each set $X_{i,j}$ is of size $2^t$ for some integer $t$; otherwise,
add dummy variables to the
sets and a clause for each $X_{i,j}$ that expresses that a non-dummy variable from the set is true.
Let $c$ be the number of clauses in $F_1$.
We now describe in a number of steps the construction of $G$.

\paragraph{Variable choice gadget}
For each set $X_{i,j}$ we have a variable choice gadget.
The gadget consists of $t$ copies of a $K_3$. Each of these $K_3$'s has one vertex marked with 0, one vertex marked with 1, and one vertex of degree two, i.e., this latter vertex has no other neighbors in the graph. 

The intuition is the following. We represent each element in $X_{i,j}$ by a unique $t$-bit bitstring (recall
that we ensured that $|X_{i,j}|=2^t$). Each $K_3$ represents one bit,
and together these bits describe one element of $X_{i,j}$, namely the variable that we set to be true.

\paragraph{Clause checking gadget}
Consider a clause $\phi$ on $X_i \cup X_{i+1}$. We perform following construction separately for each such clause, and in particular the additional vertices defined below are not shared among clauses. 
For each $j\in [1,k]$, we add $2^t+2$ additional vertices for both $X_{i,j}$ and $X_{i+1,j}$, as follows.
\begin{itemize}
    \item We create a {\em variable representing} vertex $v_x$ for each variable $x\in X_{i,j}\cup  X_{i+1,j}$.  We connect $v_x$ to the each vertex that represents a bit in the complement of the bit string representation of $x$.
    \item We add two new vertices for $X_{i,j}$ and also for $X_{i+1,j}$ (see $z_1$ and $z_2$ in Figure~\ref{figure:dslogclause}). These vertices are incident to all variable representing vertices for variables from the corresponding set.
\end{itemize}
We call this construction
a {\em variable set} gadget. 

The clause checking gadget for $\phi$ consists of $2k$ variable set gadgets (for each $j\in [1,k]$, one for each set of the
form $X_{i,j}$ and $X_{i+1,j}$), and one additional
{\em clause vertex}. The clause vertex is incident to all vertices 
in the current gadget that represent a variable that satisfies that clause. In total, we add $2 \cdot k \cdot (2^t+2) + 1$ variables per clause.

The intuition is the following. From each triangle, we place either the vertex marked with 0 or with 1 in the dominating set. This encodes a variable for each $X_{i,j}$ as follows. The vertices in the triangles dominate all but one of the variable representing vertices; the one that is not dominated
is precisely the encoded variable. This vertex is also placed in the dominating set (we need to pick at least one variable representing vertex in order to dominate the vertices $z_1$ and $z_2$ private to $X_{i,j}$). The clause is satisfied exactly when the clause vertex has a neighbor in the dominating set.

\begin{figure}
    \centering
    \includegraphics{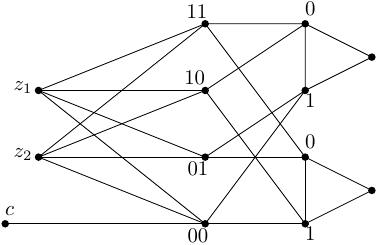}
    \caption{Example of clause gadget. Here, $t=2$ and only the variable with index 00 satisfies clause $c$}
    \label{figure:dslogclause}
\end{figure}

The graph $G$ is formed by all variable choice and clause checking gadgets, for all variables and clauses. Recall that $c$ is the number of clauses in $F_1$.

\begin{claim}
There is a dominating set in $G$ of size 
$ r k  t %%%% one vertex for each triangle in the variable choice gadgets
+
2  k c (r-1) %%%% we have c(r-1) clauses. Each on 2k sets X_{i,j} -- one vertex each
$, if and only if the
instance of {\sc Partitioned Regular Chained Weighted Positive CNF-Satisfiability} has a solution.
\end{claim}

\begin{proof}
Suppose we have a solution for the latter. Select in the variable choice gadgets the bits that encode the true
variables and put these in the dominating set $S$. As we have $rk$ sets of the form $X_{i,j}$, and each is represented by $t$ triangles of which 
we place one vertex per triangle in the dominating set, we already used $rkt$ vertices.

For each clause, we have $2k$ sets of the form $X_{i,j}$. For each, take the vertex that represents
the true variable of the set and place it in $S$. 
Note that all vertices that represent other variables are dominated by
vertices from the variable choice gadget. For each set $X_{i,j}$, the chosen variable representing vertex
dominates the vertices marked $z_1$ and $z_2$. As the clause is satisfied, the clause vertex is dominated by the vertex
that represents the variable representing vertex for the variable that satisfies it. This shows $S$ is a dominating set.
We have $(r-1)c$ clauses in total, and for each we have $2k$ sets of the form $X_{i,j}$. We placed one variable representing vertex for each of these $(r-1)c \cdot 2k$ sets in the dominating set. In total, $|S|=rkt + 2kc(r-1)$ as desired.

For the other implication, suppose there is a dominating set $S$ of size 
$ r k  t +2 k c (r-1)$.

We must contain one vertex from each triangle in a variable choice gadget (as the vertex with degree two must be dominated). Thus, these gadgets contain at least $r k t$ vertices from the
dominating set.

Each of the vertices marked $z_1$ and $z_2$ must be dominated. 
So we we must choose at least one variable choice vertex per variable set gadget. This contributes at least $2 k c (r-1)$ vertices.

It follows that we must place exactly those vertices and no more: the set $S$ contains exactly one vertex from each triangle in a variable choice gadget and exactly one variable selection vertex per variable set gadget. Note that we cannot replace a variable selection vertex by a vertex marked $z_1$ or $z_2$: either it is already dominated, or we also must choose its twin, but we do not have enough vertices for this.

From each triangle in a variable choice gadget,
we must choose either the vertex marked 0 or 1, since otherwise one of the variable selection vertices will not be dominated. This dominates all but one of the variable selection vertices, which must be the variable selection vertex that is in the dominating set. We claim that we satisfy the formula by setting exactly these variables to true and all others to false. (Note also
that we set exactly one variable per set $X_{i,j}$ to true this way.) Indeed, the clause vertex needs to be dominated by one of the variable selection vertices, and by definition this implies that the corresponding chosen variable satisfies the clause.
\end{proof}

The path decomposition can be constructed as follows. 
Let $V_i$ for $i\in [1,r-1]$ be the set of all vertices in the variable choice gadgets of
all sets $X_{i,j}$ and $X_{i+1,j}$ with $j\in [1,k]$. For any given $i$, we create a path decomposition that covers all vertices of $V_i$ and all clause gadgets connected to $V_i$. We do this by sorting the clauses in any order, and then one by one traversing the clause checking gadgets. Each time, we begin by adding the $z$-vertices and the clause vertex of the gadget to the current bag. Then, iteratively we add and remove all the variable representing vertices one by one. After this, we remove the $z$-vertices and clause vertex from the bag and continue to the next clause. At any point, there are at most $6k\cdot t + 4k + 2$ vertices in a bag: $6k\cdot t$ from the set $V_i$, $4k$ $z$-variables, one clause vertex and one variable representing vertex. We then continue the path decomposition by removing all vertices related to variable choice gadgets of the set $X_i$ and adding all vertices related to variable choice gadgets of the set $X_{i+2}$. 

As the construction can be executed in logarithmic space, the result follows.
\end{proof}

The independent set variant is defined as follows; the clique variant is defined analogously.
\begin{verse}
{\sc Log-Pathwidth Independent Set}\\
{\bf Input:} Graph $G=(V,E)$, path decomposition of $G$ of width $\ell$, integer $K$.\\
{\bf Parameter:} $\left\lceil \ell/ \log |V| \right\rceil$. \\
{\bf Question:} Does $G$ have an independent set of size at least $K$?
\end{verse}

\begin{theorem}
{\sc Log-Pathwidth Independent Set} is XNLP-complete.
\end{theorem}
\begin{proof}
As with \textsc{Log-Pathwidth Dominating Set}, the problem is in XNLP as we can use the standard dynamic programming algorithm on graphs with bounded pathwidth, but guess the table entry for each bag. 

We show XNLP-hardness with a reduction from \textsc{Partitioned Regular Chained Weighted 
Positive CNF-Satisfiability}.
The reduction is similar to the reduction to \textsc{ Log-Pathwidth Dominating Set}, but uses different gadgets specific to \textsc{Independent Set}. The clause gadget, which will be defined later, is 
a direct copy of a gadget from \cite{LokshtanovMS2018}. 

Suppose that $F_1,q,X_1,\dots,X_r,k,X_{i,j}$ ($1 \le i \le r, 1 \le j \le k$) is the given
instance of the \textsc{Partitioned Regular Chained Weighted Positive CNF-Sat\-is\-fi\-a\-bility} problem. Let $F_1 = \{C_1,...,C_m\}$ and let $|C_i|$ be the number of variables in clause $C_i$.
We assume $|C_i|$ to be even for all $i\in [1,m]$; if this number is odd, then we can add an additional copy of one of the variables to the clause. Enlarging $|X_{i,j}|$ by at most a factor of $2$ (by adding dummy variables to $X_i$), we may assume that $t=\log|X_{i,j}|$ is an integer. (Recall that $\log$ has base 2 in this paper.)

\paragraph{Variable choice gadget}
This gadget consists of $t$ copies of a $K_2$ (a single edge), with one vertex marked with a $0$ and the other with a $1$. Each element in $X_{i,j}$ can be represented by a unique $q$-bit bitstring. One $K_2$ describes one such bit and together these $t$ bits describe one element of $X_{i,j}$, which is going to be the variable that we assume to be true. We will ensure that in a maximal independent set, we need to choose exactly one vertex per copy of $K_2$, and hence always represent a variable.

\paragraph{Clause checking gadget}
The clause checking gadget is a direct copy of a gadget from \cite{LokshtanovMS2018} and is depicted in Figure \ref{figure:islogclause2}. The construction is as follows. Let $C=x_1\vee \dots \vee  x_\ell$ be a clause on $X_i\cup X_{i+1}$ (for some $i\in [1,r]$). We create two paths $p_0,\dots,p_{\ell +1}$ and $p'_1,\dots,p'_{\ell}$ and add an edge $\{p_j,p'_j\}$ for $j\in [1,\ell]$. We then add the \emph{variable representing} vertices $v_1,\dots,v_\ell$. For $j\in[1,\ell]$, the vertex $v_j$ is connected to $p_j$ and $p'_j$ and to the vertices representing the complement of the respective bits in the representation of $x_j$. This ensures that $v_j$ can be chosen if and only if the vertices chosen from the variable choice gadget represent $x_j$.

\begin{figure}
    \centering
    \includegraphics[scale = 0.8]{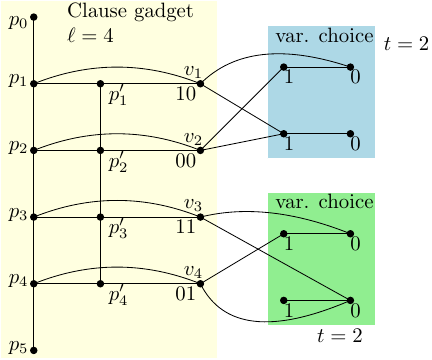}
    \caption{An example of a clause gadget is drawn with two variable gadgets. The vertex $v_1$ represents a variable which is encoded by bit string $10$, where $v_2$ represents a variable encoded by $00$. }
    \label{figure:islogclause2}
\end{figure}

For each clause, we add such a \emph{clause checking} gadget on $3\ell +2$ new vertices. The pathwidth of this gadget is $4$. 

Recall that we assume $\ell$ to be even for all clauses.
Each clause checking gadget has the property that it admits an independent set of size $\ell+2$ (which is then maximum) if and only if at least one of the $v$-type vertices is in the independent set, as proved in \cite{LokshtanovMS2018}.
Otherwise its maximum independent set is of size $\ell+1$. We try to model the clause being satisfied by whether or not the independent set contains $\ell+2$ vertices from this gadget.

The graph $G$ is formed by combining the variable choice and clause checking gadgets. 
\begin{claim}
\label{claim:logpwIS}
There is an independent set in $G$ of size 
$ rkt %%%% one vertex for each bit
+ (r-1)\sum_{i=1}^m (|C_i| +2)$ %%%% (l+2) vertices per clause checking gadget.
if and only if the
instance of {\sc Partitioned Regular Chained Weighted Positive CNF-Satisfiability} has a solution.
\end{claim}
\begin{proof}
Suppose that there is a valid assignment of the variables that satisfies the clauses. We select the bits that encode the true variables in the variable choice gadgets and put these in the independent set. This forms an independent set because only one element per variable choice gadget is set to true by a valid assignment. As we have $rk$ sets of the form $X_{i,j}$, and each is represented by $t$ times two vertices of which we place one in the independent set, we now have $rkt$ vertices in the independent set. 

For each clause checking gadget for a clause on $\ell$ variables, we add a further $\ell+2$ vertices to the independent set as follows. Let $x_i$ be a variable satisfying the clause. We add the variable representing vertex $v_i$ (corresponding to $x_i$) into the independent set. 
Since $v_i$ is chosen in the independent set, an additional $\ell+1$ vertices can be added from the corresponding clause checking gadget. In total we place $rkt + (r-1)\sum_{i=1}^m (|C_i|+2)$ vertices in the independent set. 

To prove the other direction, let $S$ be an independent set of size $rkt + (r-1)\sum_{i=1}^m (|C_i|+2)$. Any independent set can contain at most $t$ vertices from any variable choice gadget (one per $K_2$), and  at most $|C|+2$ vertices per clause checking gadget corresponding to clause $C$. Hence, $S$ contains exactly $t$ vertices per variable choice gadget and exactly $|C|+2$ vertices per clause gadget. 

Consider a variable choice gadget. Since $S$ contains exactly $t$ vertices (one per $K_2$), a variable from the set $X_{i,j}$ is encoded. It remains to show that we satisfy the formula by setting exactly these variables to true and all others to false.
Consider a clause $C$ and its related clause checking gadget. As $|C|+2$ vertices from its gadget are in $S$, this means that at least one of its vertices of the form $v_i$  is also in $S$. None of the neighbors of $v_i$ can be in $S$, so the corresponding variable $x_i$ must be set to true. Thus the clause is indeed satisfied. 
\end{proof}

A path decomposition can be constructed as follows. 
For $i=1,\dots,r$, let $V_i$ be the set of all vertices in the variable choice gadgets of the sets $X_{i,j}$ and $X_{i+1,j}$ (for all $j \in [1,k]$). 
We create a sequence of bags that contains $V_i$ as well as a constant number of vertices from clause checking gadgets.
We transverse the clause checking gadgets in any order. For a given clause checking gadget of a clause on $\ell$ variables, we first create a bag containing $V_i$ and $p_0, p_1, p'_1$ and $v_1$. Then for $s = 2,\dots,\ell$, we add $p_s, p'_s$ and $v_s$ to the bag and remove $p_{s-2},p'_{s-2}$ and $v_{s-2}$ (if these exist). At the final step, we remove $p_{\ell-1},p'_{\ell-1}$ and $v_{\ell-1}$ from the bag and add $p_{\ell+1}$. We then continue to the next clause checking gadget. 
Each bag contains at most $4k\cdot t$ from the set $V_i$ and at most $6$ from any clause gadget. 

This yields a path decomposition of width at most $4kt+6\leq 10k\log|X_{i,j}|$ which is of the form $g(k)\log(n)$ for $n$ the number of vertices of $G$ and $g(k)$ a function of the problem we reduced from, as desired. Since $G$ can be constructed in logarithmic space, the result follows. 
\end{proof}

%%%%%%%%%%%%%%%%%%%%%%%%%%%%%%%%%%%%%%%%%%%%%%

We now briefly discuss the {\sc Log-Pathwidth Clique} problem. We are given a graph $G=(V,E)$
with a path decomposition of width $\ell$, and integer $K$ and ask if there is a clique in $G$ with at least $K$ vertices.
Let $k = \left\lceil \ell /\log |V| \right\rceil $.

The problem appears to be significantly easier than the corresponding versions of {\sc Dominating Set}
and {\sc Independent Set}, mainly because of the property that for each clique $W$, there must be a bag of the path decomposition that contains all vertices of $W$ (see e.g., \cite{BodlaenderM93}.) Thus, the problem reverts to solving $O(n)$ instances of {\sc Clique} on graphs with $O(k \log n)$ vertices. 
This problem is related to a problem called {\sc Mini-Clique} where the input has a graph with a description size that is at most $k \log n$. {\sc Mini-Clique} is $M[1]$-complete under FPT Turing reductions
(see \cite[Corollary 29.5.1]{DowneyF13}), and is a subproblem of our problem, and thus
{\sc Log-Pathwidth Clique} is $M[1]$-hard. However, instances of {\sc Log-Pathwidth Clique}
can have description
sizes of $\Omega(\log^2 n)$. The result below shows that it is unlikely that {\sc Log-Pathwidth Clique} is XNLP-hard.

\begin{proposition}
{\sc Log-Pathwidth Clique} is in $W[2]$.
\end{proposition}

\begin{proof}
Suppose that we are given a graph $G=(V,E)$ and path decomposition $(X_1, \ldots, X_r)$ of $G$ of width at most $k\log n-1$. Let $K$ be a given integer for which we want to know whether $G$ has a clique of size $K$. 
We give a Boolean expression $F$ of polynomial size that can be satisfied with exactly
$2k+2$ variables set to true, if and only if $G$ has a clique of size $K$. (This shows membership in $W[2]$.)

First, we add variables $b_1, \ldots, b_r$ that are used for selecting a bag; we add the clause 
$\bigvee_{1\leq i \leq r} b_i$.

For each bag $X_i$, we build an expression that states that $G[X_i]$ has a clique with $r$ vertices as follows. Partition the vertices of $X_i$ in $k$ groups of at most $\log n$ vertices each. For each group $S\subseteq X_i$, we have a variable $c_{i,S'}$ for each subset $S'$ of vertices of the group that induces a clique in $G$. (We thus
have at most $kn$ such variables per bag.) 
The idea is that when $c_{i,S'}$ is true, then the vertices in $S'$ are part of the clique, and the other vertices in the group are not. 
The variable $c_{i,S'}$ is false if the bag $X_i$ is not selected (we add the clause $b_i \vee \neg c_{i,S'}$). 
For each group $S$, we add a clause that is the disjunction over all cliques $S'\subseteq S$
of $c_{i,S'}$ and $\neg b_i$. This encodes that if the bag $X_i$ is selected, then we need to choose a subset $S'$ from the group $S$.

With the $b$- and $c$-variables, we specified a bag and a subset of the vertices of the bag.
It remains to add clauses that verify that the total size of all selected subsets equals $K$, and that the union of all selected subsets is a clique. For the latter, we add a clause $\neg c_{i,S'} \vee \neg c_{i,S''}$ for each pair $S', S''$ of subsets of different groups for which $S' \cup S''$ does not induce a clique. 
For the former, we number the groups $1, 2, \ldots, k$, and create variables $t_{j,q}$ for
$j\in [1,k]$ and $q\in [0,K]$. The variable $t_{j,q}$ expresses that we have chosen $q$ vertices in the clique 
from the first $j$ groups of the chosen bag. For each $j \in [1,k]$, we add the clause $\bigvee_q t_{j,q}$. 
To enforce that the $t$-variables indeed give the correct clique sizes, we add a large (but
polynomial) collection of clauses. We add a variable $t_{0,0}$ that must be true (using a one-literal clause). For the $j$th group with vertex set $S\subseteq X_i$, each clique $S'\subseteq S$,
and each $q,q' \in [0,r]$ (with $q=0$ when $j=1$), whenever $q+|S'|\neq q'$, we have a clause
$\neg t_{j-1,q} \vee \neg c_{i,S'} \vee \neg t_{j,q'}$. 
Finally, we require (with a one-literal clause) that
$t_{k,K}$ holds.
If there is a satisfying assignment, then the union of all sets $S'$ for which $c_{i,S'}$ is true, forms a clique of size $K$ in the graph.

We allow $2k+2$ variables to be set to true: one variable $b_i$ to select a bag, one for $t_{0,0}$, one for a $c_{i,S'}$-variable in each of the $k$ groups and one $t_{j,q}$-variable for each $j\in [1,k]$ (where for $j=k$, $q=K$ is enforced).
\end{proof}

%%%%%%%%%% Scheduling with precedence constraints %%%%%%%%%%

\subsection{Scheduling with precedence constraints}
\label{subsection:scheduling}
In 1995, Bodlaender and Fellows~\cite{BodlaenderF95}
showed that the problem to schedule a number of jobs of unit length with precedence constraints on $K$ machines, minimizing the makespan, is $\mathrm{W}[2]$-hard, with the number of machines as parameter. A closer inspection of their proof shows that $\mathrm{W}[2]$-hardness also applies
when we take the number of machines and the width of the partial order as combined parameter.
In this subsection, we strengthen this result, showing that the problem is $\XNLP$-complete (and thus also hard for all classes $\mathrm{W}[t]$, $t\in \Z^+$.)
In the notation used in scheduling literature to characterize scheduling problems, the
problem is known as $P | prec, p_j=1 | C_{\max}$, or, equivalently, $P | prec, p_j=p | C_{\max}$.

\medskip \begin{verse}
{\sc Scheduling with Precedence Constraints}\\
{\bf Given:} $K,D$ positive integers; $T$ set of tasks; $\prec$ partial order on $T$ of width $w$.\\
{\bf Parameter:} $K+w$.\\
{\bf Question:} Is there a schedule $f:T\to [1,D]$ with $|f^{-1}\{i\}|\leq K$ for all $i\in [1,D]$ such that $t\prec t'$ implies $f(t)<f(t')$. 
\end{verse} \medskip
In other words, we parametrise $P | prec, p_j=1 | C_{max}$ by the number of machines and the width of the partial order.
\begin{theorem}
{\sc Scheduling with Precedence Constraints} is $\XNLP$-com\-plete.
\end{theorem}

\begin{proof}
To see that the problem is in $\XNLP$, define $f^{-1}\{1\},f^{-1}\{2\},\dots,$ $f^{-1}\{D\}$ in order as follows. For $i\in [0,D-1]$, we temporarily store a set $S_i\subseteq T$ containing the maximal elements of $f^{-1}[1,i]$, initialising $S_0=\emptyset$. Each such set $S_i$ forms an antichain and hence has size at most $w$, the width of the partial order on $T$. Once $S_{i-1}$ is defined for some $i\in [1,D]$, we define $f^{-1}\{i\}$ by selecting up to $K$ tasks to schedule next, taking a subset of the minimal elements of the set of `remaining tasks'
\[
R_i=\{t\in T \mid t\not \prec s \text{ for all } s\in S_{i-1}\}.
\]
By definition, for $t\in R_i$, we find that $t\not \prec s$ for all $s\in f^{-1}[1,i-1]$. In order to update $S_i$ given $S_{i-1}$, we add all selected elements (which must be maximal) and remove the elements $s\in S_{i-1}$ for which $s\prec t$ for some $t\in f^{-1}\{i\}$. After computing $S_i$ and outputting $f^{-1}\{i\}$, we remove $S_{i-1}$ and $f^{-1}\{i\}$ from our memory. At any point in time, we store at most $O((K+w)\log|T|)$ bits.

To prove hardness, we transform from the {\sc Accepting NNCCM} problem, with $k$ given counters with values in $[0,n]$ and set of checks $(s_1,\dots,s_r)$. We adapt the proof and notation of Bodlaender and Fellows \cite{BodlaenderF95}.

We create $K=2k+1$ machines and a set of tasks $T$ with a poset structure on this of width at most $3(k+1)$. Let $c=(kn+1)(n+1)$. The deadline is set to be $D= cr+n+1$. We will construct one sequence of tasks for each of the $k$ counters, with $kn+1$ repetitions of $n+1$ time slots for each of $s_1,\dots,s_r$ (one for each value in $[0,n]$ that it might like to check), and then $n$ extra time slots for `increasing' counters. Each $t\in [1,D]$ can be written in the form
\[
t= (j-1)c+\alpha(n+1)+y
\]
for some $y\in [1,n+1],~j\in [1,r],~\alpha\in [0,kn]$. We define $j(t)$ and $\alpha(t)$ to be the unique values of $j$ and $\alpha$ respectively for which this is possible. 
Our construction has the following components.
\begin{itemize}
    \item \textbf{Time sequence.} We create a sequence $a_1\prec a_2\prec \ldots \prec a_D$ of tasks that represents the time line. 
    \item \textbf{Time indicators.} By adding a task $b$ with $a_{t-1}\prec b \prec a_{t+1}$, we ensure that at time $i$ one less machine is available. We place  $k-1$ such tasks at special \emph{indicator times}. That is, we set
    \[
    I = \{(j-1) c+\alpha (n+1)+n+1\mid j\in [1,r],~\alpha\in [0,kn]\}
    \] 
    and create a task $b_t^{(x)}$ with $a_{t-1}\prec b_t^{(x)}\prec a_{t+1}$ for each $x\in [1,k-1]$ and   $t\in I$.
    \item \textbf{Counter sequences.} For each $i\in [k]$, we create a sequence $c_1^{(i)}\prec c_2^{(i)}\prec \ldots \prec c_{D-n}^{(i)}$ of tasks. If $c_{t-\ell}^{(i)}$ is planned in at the same moment as time vertex $a_{t}$ (for some $\ell\in [0,n]$ and $t\in [n+1,D]$) then this is interpreted as counter $i$ taking the value $\ell$ at that time.
    \item \textbf{Check tasks.} Let $t=(j-1)c+\alpha(n+1)+n+1\in I$ be the indicator time for $j\in [1,r]$ and $\alpha\in [0,kn]$. If $s_j=(i_1,i_2,r_1,r_2)$, then for $x \in \{1,2\}$, we add a check task $d^{(i_x)}_{t-r_x}$ `parallel' to $c^{(i_x)}_{t-r_x}$, that is, we add the precedence constraints 
    \[
    c^{(i_x)}_{t-r_x-1} \prec d^{(i_x)}_{t-r_x} \prec c^{(i_x)}_{t-r_x+1}.
    \]
\end{itemize}

\begin{figure}
    \centering
\includegraphics[width=\textwidth]{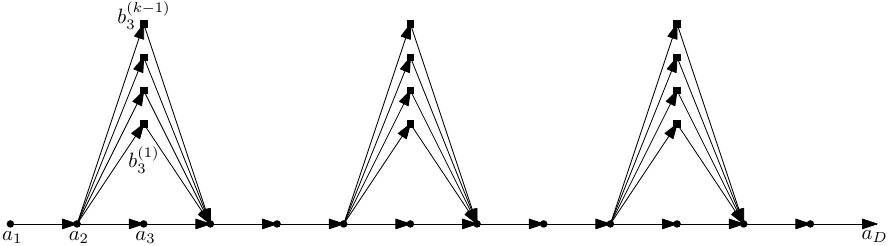}
    \caption{Illustration for a time sequence with time indicators. E.g., in the example, $3\in I$.}
    \label{figure:precedenceconstraints}
\end{figure}

Intuitively, the repetitions allow us to assume that concurrent with each time task are exactly $k$ tasks of the form $c^{(i)}_y$. For most $t\in [1,D]$, we are happy for a potential $d^{(i)}_y$ to be scheduled simultaneously (since we have $2k$ machines available). However, when $t\in I$ then we only have access to $k+1$ machines, so we may have at most one $d^{(i)}_y$ at this time; if $t$ `corresponds' to $s_j=(i_1,i_2,r_1,r_2)$, then only $d^{(i_1)}_{t-r_1},d^{(i_2)}_{t-r_2}$ can potentially be planned simultaneously with $a_t$, which happens exactly if $s_j$ rejects (due to the $i_x$th counter being at position $r_x$ for both $x\in \{1,2\}$).

The width $w$ of the defined partial order is at most $3(k+1)$. We now claim that there is a schedule if and only if the given machine accepts.

Suppose that the machine accepts. Let $(p_1,\dots,p_r)$ with $p_j=(p_j^{(1)},\dots,p^{(k)}_j)\in [0,n]^k$ be the positions of the $k$ counters before $s_j$ is checked, for $j\in [1,r]$. We schedule for $t\in [1,D],~ x\in [1,k-1],~i\in [1,k]$, when defined,
\begin{align*}
    &f(a_t)=t,\\
    &f(b_t^{x})=t, \\
    &f(c^{(i)}_{t})=t+ p_{j(t)}^{(i)}, \\
    &f(d^{(i)}_{t})=t+ p_{j(t)}^{(i)}.
\end{align*}
Recall that $j(t)$ is the unique value $j\in [1,r]$ for which we can write $t=(j-1)c+\alpha(n+1)+y$ for some $\alpha \in [0,kn]$ and $y\in [1,n+1]$. At each $t\in[1,D]$, $f^{-1}\{t\}$ contains exactly one $a_t$ and, because the counters are non-decreasing, at most one $c_{t'}^{(i)}$ and at most one $d_{t'}^{(i)}$ for each $i\in[1,k]$. So when $t\not \in I$ (and hence $b_t^x$ is not defined), $|f^{-1}\{t\}|\leq 2k+1$ as desired. 

Let $t\in I$ and $j=j(t)\in [1,r]$, $\alpha=\alpha(t)\in [0,kn]$. If $d_{t'}^{(i)}\in f^{-1}\{t\}$ for some $i\in [1,k]$, then $t'=t-p^{(i)}_j$ and $s_j=(i_1,i_2,r_1,r_2)$ with $i_x=i,~r_x=p^{(i)}_j$ for some $x\in \{1,2\}$. Since the machine accepts, there can be at most one such $i$. Hence $f^{-1}\{t\}$ contains at most one check task when $t\in I$. We again find \[
|f^{-1}\{t\}|\leq \underbrace{1}_{a_t}+\underbrace{(k-1)}_{b_t^{(x)}}+\underbrace{k}_{c_{t'}^{(i)}}+\underbrace{1}_{d_{t'}^{(i)}}\leq 2k+1.
\]

Suppose now that a schedule $f:T\to [1,D]$ exists. By the precedence constraints, we must have $f(a_t)=t$ for all $t\in [1,D]$ and $f(b_t^{x})=t$ for all $t\in I$ and $x\in [1,k-1]$. Let $j\in [1,r]$ be given. Let \[
C=\{c_t^{(i)}\mid j(t)=j,~i\in [1,k]\}.
\]
For at least one $\alpha\in [0,kn]$, it must be the case that $|f^{-1}\{t\}\cap C|=k$ for all $t$ with $\alpha(t)=\alpha$ and $j(t)=j$. This is because for each of the $k$ counter sequences, there are at most $n$ places where we may `skip'; this is why we repeated the same thing $kn+1$ times. We select the smallest such $\alpha$. Note that for all $t\in T$  with $\alpha(t)=\alpha$ and $j(t)=j$, we also find that $f(c_t^{(i)})=f(d_t^{(i)})$ for all $i\in [1,k]$.

There is a unique $t\in I$ with $\alpha(t)=\alpha$ and $j(t)=j$; for this choice of $t$, and for all $i\in [1,k]$, we set $p_j^{(i)}$ to be the value for which $c_{t-p_j^{(i)}}^{(i)}\in f^{-1}\{t\}$. This defines a vector $p_j\in [0,n]^k$ of positions which we set the counters in before $s_j$ is checked. 
Note that the values of the counters are non-decreasing, and that $s_j$ does not fail since $f^{-1}\{t\}$ contains at most one $d^{(i)}_{t'}$ (due to its size constraints), which implies that $s_j$ accepts.
\end{proof}

We remark that it is unclear if the problem is in $\XNLP$ when we take only the number of machines as parameter. In fact, it is a longstanding open problem whether {\sc Scheduling with Precedence Constraints} is $\NP$-hard when there are three machines
(see e.g., \cite{GareyJ79,ProtB18}).

%%%%%%%%% Uniform emulation

\subsection{Uniform Emulation of Weighted Paths}
\label{subsection:uepath}

In this subsection, we give a proof that the {\sc Uniform Emulation of Weighted Paths} problem is $\XNLP$-complete. The result is a stepping stone for the result that {\sc Bandwidth} is $\XNLP$-complete
even for caterpillars with hair length at most three (see the discussion in Subsection~\ref{subsection:bandwidth}).

The notion of {\em (uniform) emulation} of graphs on graphs was originally introduced by Fishburn and Finkel~\cite{FishburnF82} as a model for the simulation of computer networks on smaller computer networks. Bodlaender~\cite{Bodlaender90} studied the complexity of determining for a given graph $G$
and path $P_m$ if there is a uniform emulation of $G$ on $P_m$. In this subsection, we study a weighted
variant, and show that already determining whether there is a uniform emulation of a weighted path on a path is hard.

An \textit{emulation} of a graph $G=(V,E)$ on a graph $H=(W,F)$ is a mapping $f: V\rightarrow W$ such that
for all edges $\{v,w\}\in E$, $f(v)=f(w)$ or $\{f(v),f(w)\}\in F$. We say that an emulation is
\textit{uniform} if there is an integer $c$, such that $|f^{-1}\{w\}|=|\{v~|~ f(v)=w\}|=c$ for all $w \in W$. We call $c$ the emulation factor.

Determining whether there is a uniform emulation of a graph $H$ on a path $P_m$ is $\NP$-complete, even
for emulation factor $2$, if we allow $H$ to be disconnected. The problem to determine for a given 
\textit{connected} graph $H$ if there is a uniform emulation of $H$ on a path $P_m$ belongs to $\mathrm{XP}$ with the emulation factor $c$ as parameter \cite{Bodlaender90}.

Recently, Bodlaender~\cite{Bodlaender20} looked at the weighted variant of uniform emulation on paths, 
for the case that $H$ is a path. Now, we have a path $P_n$ and a path $P_m$, a weight function
$w: [1,n] \rightarrow \Z^+$ and ask for an emulation $f: [1,n]\rightarrow [1,m]$,
such that there is a constant $c$ with $\sum_{i\in f^{-1}\{j\}} w(i) = c$ for all $j\in [1,m]$.
Again, we call $c$ the emulation factor.

It is not hard to see that the problem, given $n$, $m$, and weight function $w: [1,n] \rightarrow \Z^+$, to determine if there is a uniform emulation of $P_n$ on $P_m$ with emulation factor $c$ is
in $\mathrm{XP}$, with the emulation factor $c$ as parameter; the dynamic programming algorithm from \cite{Bodlaender90} can easily be adapted. 

As an intermediate step for ($W[t]$)-hardness proof for {\sc Bandwidth}, Bodlaender~\cite{Bodlaender20}
showed that
{\sc Uniform Emulation of Weighted Paths} is hard for all classes $\mathrm{W}[t]$, $t\in \Z^+$.
In the current subsection, we give a stronger result, and show the same problem to be $\XNLP$-complete.
Our proof is actually simpler than the proof in ~\cite{Bodlaender20} --- by using {\sc Accepting NNCCM}
as starting problem, we avoid a number of technicalities. 

\medskip \begin{verse}
 {\sc Uniform Emulation of Weighted Paths}\\
 \textbf{Input:} Positive integers $n$, $m$, $c$, weight function $w: [1,n] \rightarrow [1,c]$.\\
 \textbf{Parameter:} $c$.\\
 \textbf{Question:} Is there a function $f: [1,n] \rightarrow [1,m]$, such that $f$ is a uniform emulation of $P_n$ on $P_m$ with emulation factor $c$, i.e., $|f(i)-f(i+1)|\leq 1$ for all $i\in [1,n-1]$ and  $\sum_{i\in f^{-1}\{j\}} w(i) = c$ for all $j\in [1,m]$?
\end{verse} \medskip

\begin{theorem}
{\sc Uniform Emulation of Weighted Paths} is $\XNLP$-\-com\-plete.
\label{theorem:pathemulation}
\end{theorem}

\begin{proof}
We first briefly sketch a variant of the dynamic programming algorithm (see \cite{Bodlaender90}) that shows membership in $\XNLP$.
For $i$ from $1$ to $m$, guess which vertices are mapped to $i$. Keep this set, and the sets for $i-1$ and $i-2$ in memory. We  reject when neighbors of vertices mapped to $i-1$ are not mapped to $\{i-2,i-1,i\}$
(or not to $\{1,2\}$ or $\{m-1,m\}$ for $i=1$ or $i=m$ respectively) or when the total weight of vertices mapped to $i$ does not equal $c$. We also check whether $cm=\sum_{i=1}^n w(i)$. Since $P_n$ is connected, and $f(i)$ is defined for some $i\in [1,n]$, we find that $f(i)$ is defined for all $i\in [1,n]$ (since our check ensures that $f(j)$ is defined for each neighbor $j$ of $i$ in $P_n$, and then also for all neighbors of $j$ etcetera). We have $O(c)$ vertices from $P_n$ in memory, and thus $O(c \log n)$ bits.

To show that the problem is $\XNLP$-hard, we use a transformation from {\sc Accepting NNCCM}.
Suppose that we are given an NNCCM$(k,n,s)$, with $s$ a sequence of $r$ checks. 
We build an instance of {\sc Uniform Emulation of Weighted Paths} as follows.

First, we define a number of constants:
\begin{itemize}
    \item $d_1 = 3k+2$,
    \item $d_2 = k \cdot d_1 +1$,
    \item $d_3 = k \cdot d_2 +1$,
    \item $c = 2k\cdot d_3 + 1$,
    \item $n_0 = 3n+1$,
    \item $M = 1+ (r+1)\cdot n_0$.
\end{itemize}

We construct a weighted path $P_N$ for some value $N$ whose value follows from our construction below, and ask for a uniform emulation of $P_N$ to $P_M$, with emulation
factor $c$. 
The path $P_N$ has three subpaths with different functions. We concatenate these in order to obtain $P_N$. \begin{itemize}
    \item The floor. We ensure that the $i$th vertex of the floor must be mapped to the $i$th vertex of $P_M$. We will use this to be able to assume that the counter components always `run from left to right'.
    \item The $k$ counter components. Each counter component models the values for one of the counters from the NNCCM and the goal is to ensure the emulation of this part is possible if and only if all the checks succeed.
    \item The filler path. This is a technical addition aimed to ensure that a total weight of $c$ gets mapped to $j$ for all $j\in [1,M]$.
\end{itemize}
The floor consists of the following $M$ (weighted) vertices.
\begin{itemize}
    \item A vertex of weight $c - k \cdot d_2$. 
    \item A path with $M-2$ vertices. The $i$th vertex of $P_N$ (and hence the $(i-1)$th of this part) has weight $c - 2d_1 +1$ if $i = n_0 \cdot j+1$ for some $j$ and weight $c - 3d_1$ otherwise. The positions of the higher weight of $c-2d_1+1$ are called {\em test positions}.
    \item A vertex of weight $c - k \cdot d_3 -1$.
\end{itemize}

We have $k$ counter components, that give after the floor, the successive parts of the `large path'.
The $q$th counter component has the following successive parts.
\begin{itemize}
    \item We start with $M-2$ vertices of weight 1. 
    \item We then have a vertex of weight $d_2$; this vertex is called the {\em left turning point}.
    \item Then we have $M-2+n$ vertices of weight either $1$ or $d_1$. The $i$th vertex has weight $d_1$ if and only if
        $i = n_0\cdot j + \alpha$ and the check $s_j$ verifies whether counter $q$ is equal to $\alpha$. The $i$th vertex has weight $1$ otherwise.
     We call this the {\em main path} of the counter component.
    \item A vertex of weight $d_3$. This vertex is called the {\em right turning point} of the counter component.
\end{itemize}
Assuming the left and right turning points are mapped to $1$ and $M$, the main path of the gadget will be mapped between $2$ and $M-1$. These main paths will tell us what the values of the counters are at any moment. The number of shifts then stands for the number a counter gadget represents, i.e. if the $i$th vertex of the main path is mapped to $i-\alpha +1$, then $\alpha$ is the value of the counter at that moment. In this way, assuming that left and right turning points are mapped to $1$ and $M$, the value of counters can only increase. 

We add a `filler path' to ensure the total weight of all vertices of $P_M$ is at least $Mc$, by adding a path of $\min\{\gamma-Mc,0\}$ vertices of weight $1$. 

\begin{claim}
The NNCCM described by $(k,n,s)$ accepts if and only if $P_N$ with the defined vertex weights has a uniform emulation on $P_M$,
\label{cl:nnccm2ue}
\end{claim}

\begin{proof}
If the NNCCM accepts, then we can build a uniform emulation as follows. Fix some accepting run and let $q(j)$ be the value of counter $q$ at the $j$th check in this run. 

Map the $i$th vertex of the floor to $i$. 
Now, successively map each counter component as follows.
We start by `moving back to 1', i.e., we map the $M-2$ vertices of weight 1 to $M-1$, $M-2$, \ldots, $2$ and map the left turning point to $1$. 
To map the vertices on the main path of this counter component, we use the following procedure. For convenience, we define $q(r+1)=n$.
\begin{itemize}
    \item Set $i=1$, $p=2$. We see $p \in [1,M]$ as a position on $P_M$, and $i \in [1,M+n]$ as the number of the vertex from the main path we are currently looking at. 
    The number $i-(p-1)$ represents the value of the counter.
    \item Set $j$ to be the smallest integer with $p\leq  n_0 \cdot j+1$; if $j\leq r$, then $n_0\cdot j+1$ is the first test position (defined in the floor gadget) after $p$.
    \item If $i-(p-1)<q(j)$, and $p$ is in the interval $[n_0 \cdot (j-1)+1+n+1,n_0 \cdot j - n]$, then we map both the $i$th and the $(i+1)$th vertex of the main path to $p$. We increase $i$ by $2$ and $p$ by $1$; the increase of $i-(p-1)$ by one represents the increase of the counter by one. Otherwise, we map the $i$th vertex of the main path to $p$, and  increase both $i$ and $p$ by one.
\end{itemize}
Since $q(r+1)=n$, in the end we will map $M-2+n$ vertices to $M-2$ positions. 

Note that the final step above does not decrease the value of $i-(p-1)$. 
The interval $[n_0 \cdot (j-1)+1+n+1, n_0 \cdot j - n]$
contains at least $n\geq  q(j)$ integers, and thus we may map two vertices to the same position until we obtain $i-(p-1) \geq q(j)$.
We then stop increasing the value $i-p+1$, so $i-p+1 = q(j)$ at the $j$th test position $p=n_0\cdot j+1$, and map a single vertex to this. 

What vertex gets mapped to the $j$th test position? If the $q$th counter does not participate in the $j$th check, then the vertex will have weight one.
Otherwise, suppose that check $s_j$ verifies whether counter $q$ is equal to the value $\alpha$.
The vertex $i$ mapped to the $j$th test position has weight $d_1$
if and only if $i=n_0\cdot j+\alpha $, which is equivalent to $q(j)=\alpha$ since $i-(p-1)=q(j)$ and $p= n_0\cdot j+1$. 
As we were considering an accepting run for the NNCCM, there is at most one counter component that maps a vertex of weight $d_1$ (rather than $1$) to the $j$th test position.

After we have placed the main path, we send the right turning point to $M$ and continue with the next counter component or, if $q=k$, the filler path.
At this point, $1$ and $M$ have received weight exactly $c$ (from  one floor vertex and $k$ turning points). For all test positions, we have a floor vertex of weight $c-2d_1+1$, at most one heavy vertex of weight $d_1$, and at most $2k$ vertices of weight one (at most two per counter component). For a vertex on $P_M$ that is not $1$, $M$ or a test position,
we have a floor vertex of weight $c-3d_1$, at most two heavy vertices of weight $d_1$ and at most $3k$
vertices of weight one (per counter component, it obtains at most two vertices for the main path and one for path going left).
In all cases, we have a weight less than $c$.

We now use the filler path to make the total weight equal to $c$ everywhere. (By definition, the filler path has enough weight 1 vertices available for this.) 

Conversely, suppose that we have a uniform emulation $f$.
We cannot map two floor vertices to the same vertex as each has a weight larger than $c/2$. As we have
$M$ floor vertices, one of the following two statements holds: 
\begin{enumerate}
    \item For all $i$, the $i$th floor vertex is mapped to $i$.
    \item For all $i$, the $i$th floor vertex is mapped to $M-i+1$.
\end{enumerate}
Without loss of generality, we assume that the $i$th floor vertex is mapped to $i$.

We first show that each right turning point is mapped to $M$ and each left turning point is mapped to $1$.
Since 
$M$ is the only vertex where the available weight (which is at most $c$ minus the weight of the floor vertex) is at least
the weight of a right turning point, all right turning points must be mapped to $M$.
After this, there is no more weight available on this vertex, and the only vertex where a left turning point can fit is $1$.

We now use the test positions to define values for the counters.
Because the main path of the counter component is from
a left turning point (mapped to 1) to a right turning point (mapped to $M$), at least one vertex of the main path is mapped to the $j$th test position $n_0\cdot j+1$. If the $i$th vertex of the main path is mapped
to the test position, then we set $q(j) = i - n_0\cdot j$. (If multiple vertices are mapped to the test position, we choose arbitrarily one of those values.) We find that the values we defined for the counters are non-decreasing since the difference between $i=i(j)$ and $n_0\cdot j$ can only become larger for larger $j$. 

We are left to prove that this defines an accepting sequence. Assume towards a contradiction that the check $s_j=(q,q',\alpha_q,\alpha_{q'})$ fails because of $q(j)=\alpha_q$ and $q'(j)=\alpha_{q'}$. That would imply both the $(n_0\cdot j + q(j))$th vertex from the $q$th counter component and the $(n_0\cdot j + q'(j))$th vertex from the $q'$th counter component were mapped to test position $n_0\cdot j+1$.
Both these vertices have weight $d_1$, by the construction of the main path of counter components. Since we already have a total weight of $c - 2d_1+1$ from the floor gadget, the total weight mapped to the $j$th test position is $>c$. This is a contradiction. Hence the given solution for the NNCCM instance is a valid one.
\end{proof}

Theorem~\ref{theorem:pathemulation} now follows from Claim~\ref{cl:nnccm2ue} and 
observing the resources (logarithmic space, fpt-time) by the transformation.
\end{proof}

In a later proof, we need the following variant of Theorem~\ref{theorem:pathemulation}.

\begin{corollary}
    {\sc Uniform Emulation of Weighted Paths} with $f(1)=M$ is $\XNLP$-\-com\-plete.
\label{corollary:pathemulation1M}
\end{corollary}

\begin{proof}
    In the proof of Theorem~\ref{theorem:pathemulation}, we showed that if there is
    a solution, then there is one with $f(1)=1$. By symmetry, we can also require
    that $f(1)=M$.
\end{proof}

%%%%%%%%%%%% Bandwidth 

\subsection{Bandwidth}
\label{subsection:bandwidth}

In this subsection, we discuss the $\XNLP$-completeness problem of the {\sc Bandwidth} problem. The question
where the parameterized complexity of {\sc Bandwidth} lies was actually the starting point for the
investigations whose outcome is reported in this paper; with the main result of this subsection (Theorem~\ref{theorem:bandwidth}) we answer a question that was asked over a quarter of a century ago.

In the {\sc Bandwidth} problem, we 
ask if the bandwidth of a given graph $G$ is 
at most $k$.
The problem models the question to permute rows and columns of a symmetric matrix, such that all
non-zero entries are at a small `band' along the main diagonal.
Already in 1976, the problem was shown to be $\NP$-complete by Papadimitriou~\cite{Papadimitriou76}. Later,
several special cases were shown to be hard; these include caterpillars with hairs of length at most three \cite{Monien86}. A \textit{caterpillar} is a tree where all vertices of degree at least three are on a common path; the \textit{hairs} are the paths attached to this main path.

We are interested in the parameterized variant of the problem, where the target bandwidth is the parameter:

\medskip \begin{verse}
{\sc Bandwidth}\\
{\bf Given}: Integer $k$, undirected graph $G=(V,E)$\\
{\bf Parameter}: $k$\\
{\bf Question}: Is there a bijection $f:V \rightarrow [1,|V|]$ such that for all edges
$\{v,w\}\in E$: $|f(v)-f(w)|\leq k$?
\end{verse} \medskip

{\sc Bandwidth} belongs to $\mathrm{XP}$. In 1980, Saxe~\cite{Saxe80} showed that {\sc Bandwidth} can be solved
in $O(n^{k+1})$ time; this was later improved to $O(n^{k})$ by Gurari and Sudborough~\cite{GurariS84}.
In 1994, Bodlaender and et.~\cite{BodlaenderFH94} reported that {\sc Bandwidth} for trees is $\mathrm{W}[t]$-hard 
for all $t\in \Z^+$ --- the proof of that fact was published 26 years later \cite{Bodlaender20}.
A sketch of the proof appears in the monograph by Downey and Fellows~\cite{DowneyF99}. More recently,
Dregi and Lokshtanov~\cite{DregiL14} showed that {\sc Bandwidth} is
$\mathrm{W}[1]$-hard for trees of pathwidth at most two. In addition, they showed that there is no
algorithm for {\sc Bandwidth} on trees of pathwidth at most two with running time 
of the form $f(k) n^{o(k)}$ assuming the 
Exponential Time Hypothesis. 

Below, we show that {\sc Bandwidth} is
XNLP-complete for caterpillars with hairs of
length at most three.\footnote{The reduction from
{\sc Uniform Emulation of Weighted Paths}
to this problem was reported in
\cite{Bodlaender20}, at WG 2020, where it was used
to show that
{\sc Bandwidth} is $\mathrm{W}[t]$-hard for all $t$.}
This reduction uses gadgets from the $\NP$-completeness proof by Monien~\cite{Monien86}.

\begin{lemma}
{\sc Bandwidth} is in $\XNLP$.
\label{lemma:bandwidthinxnlp}
\end{lemma}
\begin{proof}
This can be seen in different ways: one can look at the $\mathrm{XP}$ algorithms for {\sc Bandwidth} from
\cite{Saxe80} or \cite{GurariS84}, and observe that when instead of making full tables in the dynamic programming algorithm, we guess from each table one entry,
one obtains an algorithm in $\XNLP$.

Alternatively, we loop over all connected components $W$ of $G$, compute its size and for each, guess for $i$ from $1$ to $|W|$ the $i$th vertex in the linear ordering, i.e.,  $f^{-1}(i)$. Keep
the last $2k+1$ guessed vertices in memory, and verify that all neighbors of $f^{-1}(i-k)$ belong to
$f^{-1}[i-2k,i]$.
\end{proof}

\begin{theorem}
{\sc Bandwidth} is XNLP-complete, when restricted to caterpillars with hair length at most three.
\label{theorem:bandwidth}
\end{theorem}

\begin{proof}

To prove Theorem~\ref{theorem:bandwidth}, we transform from the {\sc Weighted Path Emulation with $f(1)=M$} problem (cf.~\ref{corollary:pathemulation1M}).

Let $P_N$ and $P_M$ be paths with weight function 
$w: \{1, \ldots, N\} \rightarrow \{1, \ldots, c\}$, and let $c = \sum_{i=1}^N w(i) / M$ is the emulation factor. (Note that we may always restrict our weights to be between $1$ and $c$ since otherwise no emulation can exist.)
Thus, $\sum_{i=1}^N w(i) = cM$. Recall that we parameterized this problem by the value $c$.

Assume that $c> 6$; otherwise, obtain an equivalent instance by multiplying all weights by 7.

Let $b= 12c+6$. Let $k= 9bc +b$. Note that $k$ is even.
We give a caterpillar $G=(V,E)$ with hair length at most three, with the property that
$P_N$ has a uniform emulation on $P_M$, if and only if $G$ has bandwidth at most $k$.

$G$ is constructed in the following way:

\begin{itemize}
    \item We have a {\em left barrier}: a vertex $p_0$ which has $2k-1$ hairs of length one, and is neighbor to $p_1$.
    \item We have a path with $5M-3$ vertices, $p_1, \ldots, p_{5M-3}$. As written above, $p_1$ is 
    adjacent to $p_0$. Each vertex of the form $5i-2$ or $5i$ ($1\leq i\leq M-1$) receives
    $2k-2b$ hairs of length one. See Figure~\ref{fig:spinepath}. We call this part the {\em floor}.
    \item Adjacent to vertex $p_{5M-3}$, we add the {\em turning point} from the proof of Monien~\cite{Monien86}. 
    We have vertices $v_a = p_{5M-3}$, $v_b$, $v_c$, $v_d$, $v_e$, $v_f$, $v_g$, which are successive vertices on
    a path. 
    I.e., we identify one vertex of the turning point ($v_a$) with the last vertex of the floor $p_{5M-3}$.
        To $v_c$, we add $\frac{3}{2} (k-2)$ hairs of length one; to $v_d$, we add $k$ hairs of length three, and to $v_f$ we add $\frac{3}{2} (k-2)$ hairs of length one. Note that this construction is identical to the one by Monien~\cite{Monien86}; vertex names are chosen to facilitate comparison with Moniens proof. See Figure~\ref{fig:turningpoint}.
    \item Add a path with $6N-5$ vertices, say $y_1, \ldots y_{6n-5}$, with $y_1$ adjacent to $v_g$. To each vertex of the form $y_{6i-5}$, add $ 9b \cdot w(i)$ hairs of length one.
    We call this part the {\em weighted path gadget}.
    \item Note that the number of vertices that we defined so far and that is not part of the turning point equals
    $2k + 5M-3 + 2(M-1)(2k-2b)+ 6N-5 + 9b \sum_{i=1}^N w_i = 5M + 4Mk - 2k - 4Mb + 4b + 9bcM$.
    Let this number be $\alpha$. One easily sees that $\alpha \leq (5M-2)k -1$.
   Add a path with $(5M-2)k-1-\alpha$ vertices and make it adjacent to $y_{6n}$.
    We call this the {\em filler} path.
\end{itemize}

\begin{figure}
    \centering
    \includegraphics[width=.95\linewidth]{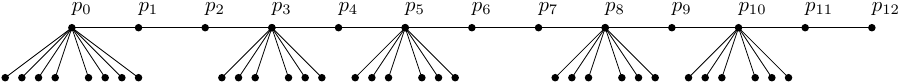}
    \caption{First part of the caterpillar: left barrier and first part of the floor.}
    \label{fig:spinepath}
\end{figure}

\begin{figure}
    \centering
    \includegraphics[width=.95\linewidth]{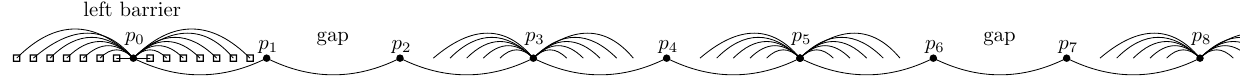}
    \caption{A layout of the left barrier and first part of the floor. The layout creates gaps where the weighted path gadgets should fit.}
    \label{fig:floorlayout}
\end{figure}

\begin{figure}[htb]
    \centering
    \includegraphics{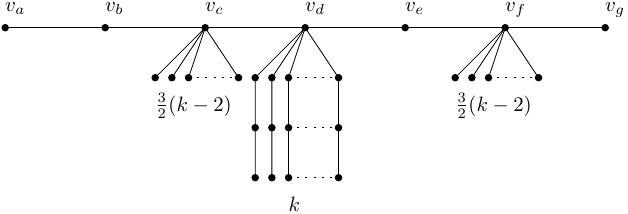}
    \caption{The Turning Point, after Monien~\cite{Monien86}}
    \label{fig:turningpoint}
\end{figure}

Let $G$ be the remaining graph. Clearly, $G$ is a caterpillar with hair length at most three. It is interesting to note that the hair lengths larger than one are only used for the turning point.

The correctness of the construction follows from the following lemma. 

\begin{lemma}
Suppose $P_N$ has a uniform emulation $g$ on $P_M$ with emulation factor $c$ with $f(1)=M$. Then the bandwidth of $G$ is at most $k$.
\label{lemma:ue2bw}
\end{lemma}

\begin{proof}
Suppose we are given a uniform emulation $f: \{1, \ldots, N \} \rightarrow \{1, \ldots, M\}$ of $P_N$ on
$P_M$ (for weight function $w$). We now can construct a layout $g$ 
of $G$ with bandwidth at most $k$ as follows.

For $1 \leq i \leq 5M-3$, set $g(p_i) = (i+1)k+1$.
Each of the hairs of $p_0$ is mapped by $g$ to
a different value in $\{1, \ldots, 2k\} \setminus \{k+1\}$. (We later define how to map the hairs of
other vertices of the floor.)

For the turning point, we set $g(v_g)=g(v_a)-1 = (5M-2)k$. The other vertices of the turning point
receive values larger than $(5M-2)k$; the remaining layout for the turning point is identical to the one described by Monien~\cite{Monien86}.

Each position of the form $(i+1)k+1 + j$ for $1\leq i\leq 5M-3$ with $j\in \{-1,+1\}$ is said
to be {\em reserved}. Note these are the positions before and after the image of the vertices $p_i$.
These positions are reserved for the filler path.

We call the positions between $g(p_{5i-4})$ and $g(p_{5i-3})$ the $i$th {\em gap}, for $1\leq i\leq M$.
Gaps $1$ till $M-1$ have at this point $k-1$ unused integers of which two are reversed; in
gap $M$ we have $k-1$ unused integers ($v_g$ is mapped in the gap) of which one is reserved.
(The gaps are used to layout `most of' the vertices of the weighted path gadget.)

We now show how to map the weighted path gadget. We start with fixing the images of vertices
of the form $y_{6i-5}$.
For each $i$, $1\leq i \leq N$, set $g(y_{6i-5})$ to be the first integer in the $f(i)$th gap 
that is so far not used and
is not reserved. As there are at most $c$ vertices of $P_N$ mapped by $f$ to a specific vertex of $P_M$,
we have that $g(y_{6i-5})$ is in the interval $[f(p_{5(i-1)})+3, f(p_{5(i-1)})+c+2]$,
thus $|g(y_{6i-5}) - g(y_{6i+1}) | \leq 5k+c$; i.e., they can be in the same, or neighboring gaps.

As $f(1)=M$, we have that $y_1$ is placed in the $M$th gap, as is, by construction its neighbor
$v_g$; thus $|g(y_1)-g(v_g)|\leq k$.

The next step is to map the vertices of the form $y_j$ with $j$ not of the form $6i-5$. We must do this
in such a way, that neighboring vertices are mapped to integers that differ at most $k$, taking into
account the mapping vertices of the form $y_{6i-5}$ as defined above. 

So, we consider, for $i \in \{1, \ldots, N-1\}$, the vertices on the path from $y_{6i-5}$ to $y_{6i+1}$ are mapped as follows.
If $y_{6i-5}$ and $y_{6i+1}$ are mapped to the same gap, then map the five vertices on this path also to this gap, to so far unused and unreserved values. 
Otherwise, define a `largest possible step to the right' from an integer $\alpha$ as the largest integer
that is at most $\alpha+k$ and is not used or reserved. If $g(y_{6i-5}) < g(y_{6i+1})$, then put
$g(y_{6i-4})$ at the largest possible step to the right from $g(y_{6i-5})$, and then 
$g(y_{6i-3})$ at the largest possible step to the right from $g(y_{6i-4})$, and continue doing so,
till we placed $g(y_{6i})$.  If $g(y_{6i+1}) < g(y_{6i-5})$, a similar construction is followed. 

\begin{myclaim}
For each $j$, $1\leq j \leq 5M-4$, we map at most $11c$ vertices of the form $y_i$ to integers in
the interval $[g(p_j), g(p_{j+1})]$.
\label{claim:used}
\end{myclaim}

\begin{proof}
If the interval is a gap, then for each such $y_i$, there is a vertex of the form $y_{6'-5}$ s at distance at most five from $y_i$ mapped to this gap. As we have at most 
$c$ vertices of the form $y_{6i'-5}$ in the gap, we have at most $11c$ vertices of the form $y_i$ mapped to this gap.

If the interval is not a gap, then for each such $y_i$, there is a vertex of the form $y_{6'-5}$ s at distance at most five from $y_i$ mapped to the last gap before the interval. Thus, we have
at most $10c$ vertices of the form $y_i$ mapped to this gap. (A more precise counting argument can give a smaller bound.)
\end{proof} 

\begin{myclaim}
For all $i$, $1\leq i < N$, $|g(y_{6i}) - g(y_{6i+1}| \leq k$.
\label{claim:4.4}
\end{myclaim}

\begin{proof}
The result clearly holds when $y_{6i-5}$ and $y_{6i+1}$ are placed in the same gap. If they are
placed in different gaps, then we consider the
case that $g(y_{6i-5}) < g(y_{6i+1})$; the other case is similar.
 
Note that each largest possible step to the right makes a jump that is at least $k - 22c-5$: from
any $k$ consecutive integers, we have used at most $22c$ for vertices of the form $y_i$ (by Claim~\ref{claim:used}, with vertices possible in two consecutive intervals), reserved four integers,
and used one for a vertex of the form $p_i$. 

Recall that $|g(y_{6i-5}) - g(y_{6i+1}) | \leq 5k+c$. Five largest jumps to the right bridge
at least $5 (k- 22c - 5)$, hence $g(6i+1)-g(6i) \leq 23c+25 < k$.
\end{proof}

The next step is to map an initial part of the filler part, to `move it' to the last number in the $M-1$st
gap. 
Suppose $f(v_N)=i$. Now, map the first vertex of the filler
path to $f(p_{5i-3})-1$, i.e., the last number in the $i$th gap (which was reserved), and then,
while we did not yet place a vertex on $f(p(5(M-1)-3))-1$ (the last number in the $M-1$st gap), map
$k$ larger than its predecessor on the path,
i.e., the second vertex goes to $g(p_{5i-3})+k-1$, the third to $g(p_{5i-3})+2k-1$, etc. In each
case, we place the vertex on the last number in the gap; these were reserved so not used by other vertices.

At this point, each open interval $(g(p_i), g(p_{i+1})$ has at most $11c+1$ used positions, and at at most
four reserved positions. Thus, at least $k-1-(11c+5) \geq k-b = 9bc$ positions are unused and not reserved
in this interval. 

For each vertex of the form $p_{5i-2}$ or $p_{5i}$ $(1\leq i \leq M-1)$, 
say $p_{i'}$
we map $k-b$ of its hairs to the interval before it: i.e., to integers in $(g(p_{i'-1}), g(p_{i'}))$ that
are unused and not reserved, and $k-b$ hairs to the interval after it, i.e., to integers in $(g(p_{i'}), g(p_{i'+1}))$ that
are unused and not reserved. 

\begin{myclaim}
For each $j$, $1\leq j\leq M$, the total number of hairs adjacent to vertices of the form $y_{6i-5}$, $1\leq i\leq N$ with $y_{6i-5}$ mapped to the $j$th gap equals $9bc$.
\label{claim:hairsingap}
\end{myclaim}
\begin{proof}
As $f$ is uniform, we have that $\sum_{i: f(i)=j} w_i = c$. A vertex $y_{6i-5}$ is placed in the $j$th gap, if and only if it contributes to this sum. As the number of hairs of $y_{6i-5}$ equals
$9b\cdot w(i)$, the result follows.
\end{proof}

For each vertex of the form $y_{6i-5}$, $1\leq i\leq N$, we map all its hairs to the gap to which it
belongs. I.e., if $f(i)=j$, then all hairs of $y_{6i-5}$ are mapped to unused and not reserved
integers in $(g(p_{5j-4},g(p_{5j-3})$. From the analysis above, we have sufficiently many such integers
to map all hairs.

Finally, we lay out the filler path to fill up all remaining unused values: go from $f(p(5(M-1)-4))-1$
to the last unused number of the $M$th gap (which is smaller than $f(p(5(M-1)-4))+k-1$, and then,
while there are unused numbers left, place each next vertex of the filler gap on the largest so far unused number.
Thus, from this point on, the filler gap goes from right to left, skipping only used numbers. 
The reserved numbers of the form $(i+1)k+1$ guarantee that by doing so, we never skip more than $k-1$ numbers, thus preserving the bandwidth condition.

This finishes the construction of the linear ordering $g$. From the analysis above, it follows that $g$
has bandwidth $k$.
\end{proof}

\begin{lemma}
Suppose the bandwidth of $G$ is at most $k$. Then $P_N$ has a uniform emulation on $P_M$ with emulation factor $c$ with $f(1)=M$.
\label{lemma:uetobw}
\end{lemma}

\begin{proof}
Suppose we have a linear ordering $g$ of $G$ of bandwidth at most $k$.

First we note that all vertices outside the left barrier and the turning point have to be mapped between
the left barrier and the turning point. For the left barrier, this is easy to see: $p_0$ has $2k$ neighbors, which will occupy all $2k$ positions with distance at most $k$ to $g(p_0)$, and thus all other vertices
must be at the same side of $p_0$ as $p_1$. 

Now, we look at the turning point, and use a result by Monien~\cite[Lemma 1]{Monien86}.

\begin{lemma}[Monien~\cite{Monien86}]
Let $g$ be a linear ordering of a graph $G$ containing the turning point as subgraph.
$|g(v_a)-g(v_g)|\leq 1$, and one of the following two cases holds.
\begin{enumerate}
    \item All vertices in the turning point except $v_a$ and $v_g$ have an image under $g$ that is
    larger than $\max\{g(v_a), g(v_g)\}$; all vertices in $G$ that do not belong to the turning point have an image that is smaller than $\min\{g(v_a), g(v_g)\}$.
        \item All vertices in the turning point except $v_a$ and $v_g$ have an image under $g$ that is
    smaller than $\min\{g(v_a), g(v_g)\}$; all vertices in $G$ that do not belong to the turning point have an image that is larger than $\max\{g(v_a), g(v_g)\}$.
\end{enumerate}
\end{lemma}

I.e., we have that $v_a$ and $v_g$ are next to each other; all other vertices of the
turning point are at one side of this pair, and all vertices not in the turning point at the other side.
Without loss of generality, suppose $g(p_0) < g(v_a)$. 

Thus, from left to right we have:
$k$ hairs of $p_0$; $p_0$; $k-1$ hairs of $p_0$; $p_1$; all vertices except $p_1$, vertices in the left barrier, and vertices in the turning point; $v_a = p_{5M-3}$ and $v_g$ (in this order, or reversed); all other vertices of the turning point.

\begin{myclaim}
$g(p_0)=k+1$, and $\{g(v_a),g(v_g)\} = \{(5M-2)k, (5K-2)k+1\}$.
\end{myclaim}

\begin{proof}
$p_0$ is placed after $k$ hairs, so must have image $k+1$.
The total number of vertices not in the turning point equals $(5M-2)k-1$, see the construction
of the filler path. The pair $\{g(v_a),g(v_g)\}$ comes after all vertices not in the turning point,
and before all other vertices from the turning point, so must go to $(5M-2)k$ and $(5M-2)k+1$.
\end{proof}

\begin{myclaim}
For all $i$, $0\leq i < 5M-3$, 
either $g(p_{i+1}) = g(p_i) +k$ or $g(p_{i+1}) = g(p_i) +k-1$. 
\end{myclaim}

\begin{proof}
The path from $p_0$ to $v_a=p_{5M-3}$ has $M-3$ edges. For each pair $p_i$, $p_{i+1}$, the difference
of the images is at most $k$ (by the bandwidth condition), but the total of the differences over all pairs on the path must
be at least $(M-3)\cdot k -1$.
\end{proof}

Now, call the positions between $g(p_{5i-5})$ AND $g(p_{5i-2})$ the $i$th {\em enlarged gap}, for $1\leq i \leq M-1$; and between $g(5M-5)$ and $g(p_{5M-3})$ the $M$th enlarged gap. 

\begin{myclaim}
Each vertex of the form $y_{6j-5}$ is mapped to an enlarged gap.
\label{claim:gap1}
\end{myclaim}

\begin{proof}
Suppose not. As $g(y_{6j-5}) > g(p_1)$ and $g(y_{6j-5}) < g_{5M-3}$, by construction of the left barrier
and turning point, we have that there is an $i$ with $g(y_{6j-5}) \in [g(p_{5i-2}), g(p_{5i})]$. 
Now, $[g(p_{5i-3}), g(p_{5i+1})]$ contains $4k-4b$ hairs attached to vertices $p_{5i-2}$ and $p_{5i}$
and $9b\cdot w(j) \geq 9b$ hairs attached to $y_{6j-5}$. The interval has size at most $4k$, but has
$4k+b$ hairs mapped to it; contradiction.
\end{proof}

Now, let $f:\{1,\ldots, N\} \rightarrow \{1, \ldots, M\}$, such that $f(j)=i$, if $y_{6j-5}$
is mapped to the $i$th enlarged gap. By Claim~\ref{claim:gap1}, $f$ is well defined.

\begin{myclaim}
$f$ is an emulation.
\label{claim:gap2}
\end{myclaim}

\begin{proof}
Note that there is a path with six edges from a vertex $y_{6j-5}$ to $y_{6(j+1)-5}$. As the distance
between vertices $p_i$ in the linear ordering is either $k$ or $k-1$, each edge in this path can
jump over at most one vertex of the form $p_\alpha$. To go from the $i$th enlarged gap to the $i+2$nd enlarged gap,
we must jump at least seven vertices of the form $p_\alpha$, so $y_{6j-5}$ and $y_{6(j+1)-5}$
are mapped to the same or neighboring enlarged gaps, thus $|f(j)-f(j+1)|\leq 1$.
\end{proof}

\begin{myclaim}
$f$ is uniform.
\label{claim:gap3}
\end{myclaim}

\begin{proof}
We show that for each $i$, $\sum_{j: f(j)=i} w(j) \leq c$. As $f$ is uniform, we have
that for each $i$: $\sum_{j: f(j)=i} w(j) = c$. 

We only give the proof for $1 < i < M$; the cases $i=1$ and $i=M$ are similar (using that no vertices
can be mapped to values used by the left blockade and turning point) and omitted.

Consider an $i$, $1<i<M$. 
For each $j$ with $f(i)=j$, all $9b \cdot w(j)$ hairs of $y_{6j-5}$ are mapped to the interval
$[g(p_{5i-6}), g(p_{5i-1})]$. This interval has size at most $5k$, contains $4k-4b$ hairs of
$p_\alpha$-vertices, thus can contain at most $8k+4b$ hairs of vertices $y_\beta$. All hairs of
vertices $y_{6j-5}$ with $f(j)=i$ must be mapped to this interval. As the number of these hairs
equals $9b$ times the sum of the weights of these vertices, we have $\sum_{j: f(j)=i} w(j) \leq c$.
\end{proof}

Finally, we observe that $f(1)=M$: $y_1$ is incident to $v_g$, and thus $y_1$ must be placed in
the $M$th enlarged gap.

Now, with this last observation, Claims~\ref{claim:gap1}, \ref{claim:gap2} and \ref{claim:gap3} together show Lemma~\ref{lemma:uetobw}.
\end{proof}

An illustration of the construction of a linear ordering, given a uniform emulation is given in Figure~\ref{figure:examplebw}.

\begin{figure}[htb]
    \centering
    \includegraphics[width= \textwidth]{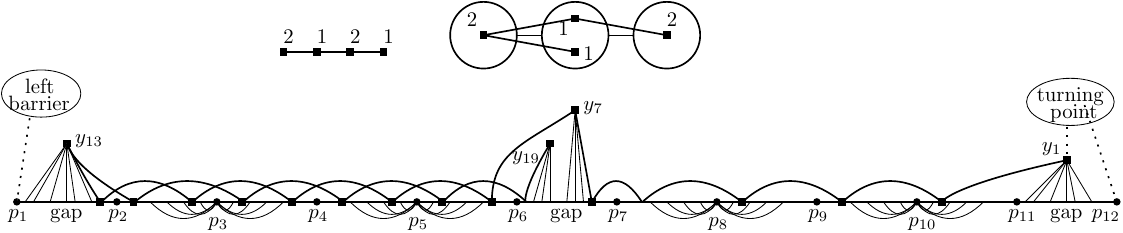}
    \caption{Illustration of part of the construction. Shown are $P_4$ with successive vertex weights 2, 1, 2, 1; a uniform emulation on $P_3$ with emulation factor 2; a layout of a part of $G$.}
    \label{figure:examplebw}
\end{figure}

As the construction of the caterpillar $G$ can be done in polynomial time and
logarithmic space, given $M$, $N$ and $w$,
Theorem~\ref{theorem:bandwidth} now follows.
\end{proof}

\subsection{Acyclic Finite State Automata Intersection}
\label{section:fsaintersection}
In the {\sc Finite State Automata Intersection} problem, we are given $k$ deterministic finite state automata
on an alphabet $\Sigma$ and ask if there is a string $s\in \Sigma^\ast$ that is accepted by each of the automata. 

In the overview of the parameterized complexity of various problems in \cite{DowneyF99}, the problem is
mentioned to be hard for all classes $\mathrm{W}[t]$, $t\in \Z^+$, when either parameterized by the number
of machines $k$ or by the combination of the number of machines
$k$ and the size of the alphabet $\Sigma$; the result is
due to Hallett, but has not been published.

More recently,
the problem and many variations were studied by Wehar~\cite{Wehar16}. Amongst others, he showed that {\sc Finite State Automata Intersection} with the number of machines as parameter is $\mathrm{XNL}$-complete. He also considered the variant where the automata are
acyclic. 
\medskip \begin{verse}
{\sc Acyclic Finite State Automata Intersection}\\
{\bf Input:} $k$ deterministic finite state automata on an alphabet $\Sigma$ for which the underlying graphs are acyclic (except for self-loops at an accepting or rejecting state). \\
{\bf Parameter:} $k$. \\
{\bf Question:} Is there a string $s\in \Sigma^\ast$ that is accepted by each of the automata?
\end{verse} \medskip

Wehar~\cite[Chapter 5]{Wehar16} showed that {\sc Acyclic Finite State Automata Intersection}
is equivalent under LBL-reductions (parameterized reductions that do not change the parameter) to a version of {\sc Timed CNTMC} (see Section \ref{subsec:previousknown}) where the given time bound $T$ is linear.
The proof technique of Wehar~\cite{Wehar16} can also be used to show that 
{\sc Acyclic Finite State Automata Intersection} is $\XNLP$-complete. We use a different, simple reduction from {\sc Longest Common Subsequence}.
\begin{theorem}
\label{thm:acyclicfsa}
{\sc Acyclic Finite State Automata Intersection} is $\XNLP$-complete.
\end{theorem}

\begin{proof}
For $\XNLP$-membership of {\sc Acyclic Finite State Automata Intersection}, the argument
from \cite[Proposition 5.1]{Wehar16} can be used: guess the solution string one character at the time, and keep track of the states of the machines during these guesses. To see this does not take too much time, let $n$ be the largest number of vertices in a graph underlying one of the machines. Since the graphs are acyclic, all machines enter an accepting or rejecting state within $n$ transitions.

The simple transformation from {\sc Longest Common Subsequence} parameterized by the number of strings $k$ to {\sc Acyclic Finite State Automata Intersection}
is given below. 

Suppose that we are given $k$ strings $s^1, \ldots, s^k \in \Sigma^\ast$ and an integer $m$. 
We consider $k+1$ automata on alphabet $\Sigma$: one for each string and one that checks the length of the solution.

The \emph{length automaton} has $m+1$ states $q_0, \ldots, q_m$, where $q_0$ is the starting state and $q_m$ is the accepting state. For each $i\in [0,m-1]$, and each $\sigma\in \Sigma$ we have a transition from
$q_i$ to $q_{i+1}$ labeled with $\sigma$. We also have for each $\sigma\in \Sigma$ a transition (self-loop) from $q_{m}$ to $q_m$.
The following claim is easy to observe.

\begin{claim}
The length automaton accepts a string $s\in \Sigma^\ast$ if and only if the length of $s$ is at least $m$.
\end{claim}

For each $i\in [1,k]$, we have a \emph{subsequence automaton} for string $s^i$.
Suppose that its length is $|s^i|=t$. The string automaton for $s^i$ has the following $t+2$ states: $q_0, \ldots, q_t, q_R$. All states except for the rejecting state $q_R$ are accepting states, and $q_0$ is the starting state.
The automaton for $s^i$ has the following transitions.
For each $j\in[0,t-1]$ and $\sigma\in \Sigma$, if the substring $s^i_{j+1} \cdots s^i_t$
contains the symbol
$\sigma$, then let $s^i_{j'}$ be the first occurrence of $\sigma$ in this substring, and take a transition from $q_j$ to $q_{j'}$ labeled with $\sigma$.
(Recall that for each state $q$ and each $\sigma\in \Sigma$, if no transition out of $q$ labeled with $\sigma$ was defined in the previous step, then we take a transition from $q$ to $q_R$ labeled with $\sigma$.)
The name `subsequence automaton' is explained by the following claim.
\begin{claim}
The subsequence automaton for string $s^i$ accepts a string $s\in \Sigma^*$ if and only if $s$ is
a subsequence of $s^i$.
\end{claim}
\begin{proof}
Note that after reading a character, the automaton moves to the index of the next occurrence of this character after the current index. So, when we read a subsequence $s$, we are in state $q_j$ when $j$ is the smallest integer with $s$ a subsequence of the substring $s^i_1 \ldots s^i_j$. When there is no such next character, then $s$ is not a subsequence and we move to the rejecting state.
\end{proof}

From the claims above, it follows that a string $s\in \Sigma^\ast$ is a subsequence of $s^1, \ldots, s^k$  of length at least $m$
if and only if $s$ is accepted by both the length automaton and the subsequence automata of $s^1, \ldots, s^k$. We can compute the automata in $O(f(k)+\log n)$ space, given the strings $s^1, \ldots, s^k$ and $m$ (with $n$ the number of bits needed to describe these).
\end{proof}

It is also not difficult to obtain $\XNLP$-completeness for the restriction to a binary alphabet.
First, add dummy symbols to $\Sigma$ to ensure that the size of the alphabet $\Sigma$ is a power of two. Each transition labeled with a dummy symbol leads to the rejecting state $q_R$, i.e., we accept the same set of strings.
Now, encode each element of $\Sigma$ with a unique string in $\{0,1\}^{\log |\Sigma|}$.
For each state $q$ in the automata, replace the outgoing transitions by a complete binary tree
of depth $\log |\Sigma|$ with the left branches labeled by 0 and the right branches labeled by 1. For a leaf of this tree, look at the string $z \in \{0,1\}^{\log |\Sigma|}$ formed by the labels when one follows the path from $q$ to this leaf. This codes a character in $\Sigma$; now let this leaf have a transition to the state reached from $q$ when this character is read. 
This straightforward transformation gives an automaton that precisely accepts the strings when we
replace each character by its code in $\{0,1\}^{\log| \Sigma|}$. If we apply the same transformation
to all automata, we obtain an equivalent instance but with $\Sigma=\{0,1\}$.
We can conclude the following result.

\begin{corollary}
{\sc Acyclic Finite State Automata Intersection} is \linebreak $\XNLP$-complete for automata with a binary alphabet.
\end{corollary}

%%%%%%% Conclusions %%%%%%%%%

\section{Conclusion}
\label{section:conclusions}
We end the paper with some discussions and open problems. We start by discussing a conjecture  on the space usage of $\XNLP$-hard problems, then discuss the type of reductions we use, and then give a number of open problems.

\subsection{Space efficiency of $\XNLP$-hard problems}
\label{section:spaceefficiency}
Pilipczuk and Wrochna~\cite{PilipczukW18} made the following conjecture. In the {\sc Longest Common Subsequence} problem, we are given $k$ strings $s^1, \ldots, s^k$ over an alphabet $\Sigma$ and an integer
$r$ and ask if there is a string $t$ of length $r$ that is a subsequence of each $s^i$, $i\in [1,k]$. 

\begin{conjecture}[Pilipczuk and Wrochna~\cite{PilipczukW18}]
The {\sc Longest Common Subsequence} problem has no algorithm that runs in $n^{f(k)}$ time and $f(k) n^c$ space, for a computable function $f$ and constant $c$, with $k$ the number of strings, and $n$ the total input size.
\label{conjecture:PW}
\end{conjecture}

Interestingly, 
this conjecture leads to similar conjectures for a large collection of problems. As {\sc Longest Common Subsequence} with the number of strings $k$ as parameter is $\XNLP$-complete~\cite{ElberfeldST15}, 
Conjecture~\ref{conjecture:PW} is equivalent to the following conjecture,
which is currently known as the \emph{Slice-wise Polynomial Space Conjecture}.
Recall that the class $\XNLP$ is the same as the class
$\operatorname{N} [ f \poly, f \log]$.

\begin{conjecture}
$\operatorname{N} [ f \poly, f \log] \not\subseteq \operatorname{D}[ n^f, f \poly]$.
\label{conjecture:swps}
\end{conjecture}

If Conjecture~\ref{conjecture:PW} holds, then no $\XNLP$-hard problem has an algorithm that uses XP time and
simultaneously `FPT' space (i.e., space bounded by the product of a computable function of the parameter
and a polynomial of the input size). Thus, $\XNLP$-hardness proofs yield conjectures about the space usage of 
$\mathrm{XP}$ algorithms, and Conjecture~\ref{conjecture:PW} is equivalent to the same conjecture for {\sc Bandwidth}, {\sc List Coloring} parameterized by pathwidth, {\sc Chained CNF-Satisfiability}, etc.

In particular, the conjecture suggests that there is a constant $k^*$ for which any deterministic Turing machine needs $\omega(\log n)$ space in order to solve \textsc{List Coloring} for $n$-vertex graphs of treewidth $k^*$; for trees the problem can still be solved in logarithmic space \cite{treeslogspace}, i.e. $k^*$ needs to be chosen strictly greater than one, but it's unclear what space efficiency can be achieved for treewidth 2, for example.\footnote{In a subtle way,
the conjecture actually does not rule out that there
is an $O(\log n)$ space algorithm for \textsc{List Colouring} for treewidth $k$ graphs, for each $k$, but with the constant hidden in the $O$-notation growing faster than any computable function $f$.}

%%% DIFFERENT REDUCTIONS
%%%
\subsection{Reductions}
\label{section:conclusion-reductions}
In this paper, we mainly used parameterized logspace reductions (pl-reductions),
i.e., parameterized reductions that
run in $f(k) + O(\log n)$ space, with $f$ a computable function.

Elberfeld et al~\cite{ElberfeldST15} use a stronger form of reductions, namely {\em parameterized first-order reductions} or pFO-reductions, where the reduction can be computed by a logarithmic time-uniform para $AC^O$-circuit family.
In~\cite{ElberfeldST15}, it is shown that {\sc Timed Non-Deterministic Cellular Automaton} and
{\sc Longest Common Subsequence} (with the number of strings as parameter) are $\XNLP$-complete under pFO-reductions. 
We have chosen to use the easier to handle notion of logspace reductions throughout the paper, and not
to distinguish which steps can be done with pl-reductions and which not.

One might want to use the least restricted form of reductions, under which $\XNLP$ remains closed, and that
are transitive, in order to be able to show hardness for $\XNLP$ for as many problems as possible. 
Instead of using $O(f(k)+\log n)$ space, one may want to use $O(f(k) \cdot \log n)$ space --- thus allowing to use a number of counters and pointers that depends on the parameter, instead of being bounded by a fixed
constant. However, it is not clear that $\XNLP$ is closed under parameterized reductions with a $O(f(k) \cdot \log n)$ space bound, as the reduction may use $O(n^{f(k)})$ time. 

To remedy this, we can simultaneously
bound the time and space of the reduction.
A {\em parameterized tractable logspace reduction} (ptl-reductions) is a parameterized reduction that simultenously uses
$O(f(k) \cdot \log n)$ space and $O(g(k) \cdot n^c)$ time, with $f$ and $g$ computable functions,
$k$ the parameter, and $n$ the input size.
One can observe that the same argument (`repeatedly recomputing input bits when needed') that shows transitivity of L-reductions (see \cite[Lemma 4.15]{AroraBarak}) can be used to show transitivity of parameterized tractable logspace reductions (and of parameterized logspace reductions). 

We currently
are unaware of a problem where we would use ptl-reductions instead of pl-reductions.
However, the situation reminds of a phenomenon that also shows up for hardness proofs for classes in the $\mathrm{W}$-hierarchy. Pl-reductions allow us to use time that grows faster than polynomial in the parameter value.
If we have an fpt-reduction that uses $O(f(k)n^c)$ time with $c$ a constant, and $f$ {\em a polynomial
function}, then this reduction is also a many-to-one reduction, and could be used in an $\NP$-hardness proof
for the unparameterized version of the problem. Most {\em but not all} fpt-reductions from the literature
have such a polynomial time bound. However, in the published 
hardness proofs, the distinction is usually not made explicit.

\subsection{Candidate $\XNLP$-complete problems}
In this paper, we showed that
the class $\XNLP$ captures the
complexity of various parameterized problems.
The common denominator of such problems appears to be
that there is some form of `linear structure' in
the problem statement: either we search for some
linear structure, or the input is linearly structured.
Recent work, following our investigations, revealed
that there are more such linear structured problems
to be complete for $\XNLP$, e.g.~\cite{BodlaenderGJJL22}; we expect that there
are much more examples. A number of candidates
are linear graph ordering problems, like
\textsc{Colored Cutwidth}, \textsc{Feasible Register
Allocation} (see \cite{BodlaenderFHWW00}), 
\textsc{Shortest Common Supersequence} as mentioned in \cite{DowneyF99},  {\sc Restricted Completion to a Proper Interval Graph with Bound\-ed Clique Size},
(see \cite{KaplanS96}), or problems known to be in XP when
parameterized by a linear width measure (like pathwidth, linear cliquewidth).
Jansen et al.~\cite{JansenKKLMS23} mentioned as open problem XNLP-hardness for two planarity testing
problems with treewidth as parameter. Bakkane and Jaffke \cite{BakkaneJaffkeIPEC22} suggested that the generalized dominating set problems which they prove to be W[1]-hard are likely to be XNLP-hard.

\section*{Acknowledgement}
The authors would like to thank the referees for useful comments on an earlier conference version of this paper.

\bibliographystyle{abbrvurl}
\bibliography{references}

\end{document}